# Complexity of Judgment Aggregation

**Ulle Endriss**                                           ULLE.ENDRISS@UVA.NL
**Umberto Grandi**                                     UMBERTO.UNI@GMAIL.COM
**Daniele Porello**                                DANIELEPORELLO@GMAIL.COM
*Institute for Logic, Language and Computation*
*University of Amsterdam*
*Postbus 94242, 1090 GE Amsterdam*
*The Netherlands*

## Abstract

We analyse the computational complexity of three problems in judgment aggregation:
(1) computing a collective judgment from a profile of individual judgments (the winner
determination problem); (2) deciding whether a given agent can influence the outcome
of a judgment aggregation procedure in her favour by reporting insincere judgments (the
strategic manipulation problem); and (3) deciding whether a given judgment aggregation
scenario is guaranteed to result in a logically consistent outcome, independently from what
the judgments supplied by the individuals are (the problem of the safety of the agenda).
We provide results both for specific aggregation procedures (the quota rules, the premise-
based procedure, and a distance-based procedure) and for classes of aggregation procedures
characterised in terms of fundamental axioms.

## 1. Introduction

Judgment aggregation (JA) is a branch of social choice theory that studies the properties
of procedures for amalgamating several agents' individual judgments regarding the truth or
falsity of a set of inter-related propositions into a collective judgment reflecting the views of
that group of agents as a whole (List & Pettit, 2002; List & Puppe, 2009). A by now classic
example is due to Kornhauser and Sager (1993): Suppose three judges have to decide on a
legal case involving a possible breach of contract. Two relevant propositions are that there
really has been a binding contract rather than just an informal promise (proposition $p$) and
that the defendant broke her promise (proposition $q$). The defendant should be pronounced
guilty if the conjunction of these two propositions is found to be true ($p \wedge q$). Our judges
take the following views on the matter:

|          | $p$ | $q$ | $p \wedge q$ |
|----------|-----|-----|-----|
| Judge 1  | Yes | Yes | Yes |
| Judge 2  | Yes | No  | No  |
| Judge 3  | No  | Yes | No  |
| Majority | Yes | Yes | No  |

Note that the position of each individual judge is logically consistent. However, if we
aggregate this information using the majority rule (i.e., if we accept a proposition if and
only if a strict majority of the judges do), then we arrive at a collective judgment set that
is inconsistent. This paradoxical outcome, variations of which are known as the *doctrinal*

*paradox* (Kornhauser & Sager, 1993) or the *discursive dilemma* (Pettit, 2001), has inspired an important and fast growing literature on JA, starting with the seminal contribution of List and Pettit (2002), who showed that in fact no aggregation procedure satisfying certain axioms encoding natural desiderata can avoid this kind of paradox.

The literature on JA has largely developed in outlets associated with Philosophy, Economic Theory, Political Science, and Logic, but recently JA has also come to be recognised as being relevant to AI, particularly to the design and analysis of multiagent systems. The reasons are clear: in a multiagent system, different autonomous software agents may have different "opinions" on the same issues (maybe due to a difference in access to the relevant information, or due to different reasoning capabilities), and some joint course of action needs to be extracted from these diverse views. Indeed, in AI, the related problem of *belief merging* has been studied for some time (see, e.g., Konieczny & Pino Pérez, 2002; Maynard-Zhang & Lehmann, 2003; Chopra, Ghose, & Meyer, 2006; Everaere, Konieczny, & Marquis, 2007), and there are interesting parallels between that literature and JA (Pigozzi, 2006). JA has also been found to be relevant to the analysis of abstract argumentation frameworks widely studied in AI (Caminada & Pigozzi, 2011; Rahwan & Tohmé, 2010).

Given the relevance of JA to AI, it is important to understand its computational aspects. However, to date, these have only received relatively little attention in the literature. This can of course be explained by the origins of the field in Law, Economics, and Philosophy. In other domains of social choice, particularly voting and fair division, on the other hand, the recent focus on computational aspects has been very successful and has given rise to the field of computational social choice (Chevaleyre, Endriss, Lang, & Maudet, 2007; Brandt, Conitzer, & Endriss, 2012).

To help bridge this gap, in this paper we shall analyse the computational complexity of three important problems that arise in JA:

- **Winner determination.** The *winner determination problem* is the problem of computing the result of applying a given aggregation procedure to a given profile of individual judgment sets. It is of immediate practical relevance to all applications of JA. We obtain both positive and negative results: for two types of aggregation procedures, namely the *quota rules* and the *premise-based procedure*, the winner determination problem is easily seen to be polynomial, while for a certain *distance-based procedure* we obtain an interesting intractability result, establishing completeness for parallel access to NP that mirrors corresponding results in voting theory for the Dodgson rule (Hemaspaandra, Hemaspaandra, & Rothe, 1997), the Young rule (Rothe, Spakowski, & Vogel, 2003) and the Kemeny rule (Hemaspaandra, Spakowski, & Vogel, 2005).

- **Strategic manipulation.** An agent may try to influence the result of aggregation in her favour by reporting a set of judgments that is different from her truthfully held beliefs. The *manipulation problem* asks, for a given aggregation procedure, a given profile of judgment sets, and a given agent, whether that agent has the opportunity to manipulate successfully in this situation. For one natural way of defining preferences on top of the JA framework (namely in terms of the *Hamming distance*) and for aggregation procedures that are *independent* and *monotonic*, it is well-known that agents will never have an incentive to manipulate (Dietrich & List, 2007c). In other cases, it is interesting to explore how hard it is to solve the manipulation problem, as high complexity might signal a certain level of immunity against manipulation. In the context of voting, this kind of question has lead to a series of interesting and important results (Bartholdi, Tovey, & Trick, 1989; Faliszewski & Procaccia, 2010), even if we have to be careful not to over-interpret theoretical intractability results as necessarily providing protection in practice (Walsh, 2011). For one widely used procedure (with an easy winner determination problem), namely the *premise-based procedure*, we are able to show NP-completeness for the manipulation problem.

- **Safety of the agenda.** The paradox presented above shows that for some aggregation procedures it is possible to obtain a collective judgment set that is logically inconsistent, even though each of the judgment sets supplied by the individuals is consistent. An important parameter determining the possibility of such a paradox is the *agenda*, the set of propositions on which to pass judgment. For a given aggregation procedure, the *problem of the safety of the agenda* asks whether a given agenda is safe in the sense that no profile of individual judgment sets that are consistent can ever result in a collective judgment set that is inconsistent. For various classes of aggregation procedures, defined in terms of classical *axioms*, we prove *safety theorems* that fully characterise agendas that are safe in this sense and we relate our results to known *possibility theorems* from the JA literature (List & Puppe, 2009). We then study the complexity of deciding whether a given agenda meets the safety conditions identified and we find that this is typically a highly intractable problem located at the second level of the polynomial hierarchy.

These results build on and extend our earlier work on the complexity of judgment aggregation (Endriss, Grandi, & Porello, 2010a, 2010b).

The remainder of this paper is organised as follows. In Section 2 we introduce the formal framework of JA, including several concrete aggregation procedures and the most important axioms used to define desiderata for such procedures. Section 2 also includes proofs of some simple representation results that characterise aggregation procedures that satisfy certain combinations of these axioms. Section 3 is devoted to the study of the complexity of the winner determination problem and Section 4 does the same for the manipulation problem. In Section 5 we then introduce the problem of the safety of the agenda, prove several agenda characterisation theorems establishing necessary and sufficient conditions for safety, and finally study the complexity of deciding whether those conditions are satisfied. Section 6 reviews related work on computational aspects of JA and Section 7 concludes with a discussion of possible avenues for future work.

Throughout this paper, we shall assume familiarity with the basics of complexity theory up to the notion of NP-completeness. Helpful introductions include the textbooks by Papadimitriou (1994) and Arora and Barak (2009).

## 2. The Formal Framework of Judgment Aggregation

In this section we provide a succinct exposition of the formal framework of JA (List & Puppe, 2009), which originally was laid down by List and Pettit (2002) and since then has been further refined by a number of authors, notably Dietrich (2007). We also define three concrete (families of) aggregation procedures and we discuss the most important axiomatic properties from the literature. Finally, we prove a number of representation results, which have the status of *folk theorems* in the JA literature and often play a crucial role in the proofs of more complex results, but which have rarely been stated explicitly.

## 2.1 Notation and Terminology

Let $\mathcal{L}$ be a set of propositional formulas built from a finite set of propositional variables using the usual connectives $\neg, \wedge, \vee, \rightarrow, \leftrightarrow$, and the constants $\top$ ("true") and $\bot$ ("false"). For every formula $\alpha$, define $\sim\alpha$ to be the *complement* of $\alpha$, i.e., $\sim\alpha = \neg\alpha$ if $\alpha$ is not negated, and $\sim\alpha = \beta$ if $\alpha = \neg\beta$ for some formula $\beta$. We say that a set $\Phi$ is *closed under complementation* if it is the case that $\sim\alpha \in \Phi$ whenever $\alpha \in \Phi$.

**Definition 1.** *An **agenda** is a finite nonempty set $\Phi \subseteq \mathcal{L}$ that does not contain any doubly-negated formulas and that is closed under complementation.*

That is, in a slight departure from the common definition in the literature (List & Puppe, 2009), we allow for *tautologies* and *contradictions* in the agenda. The reason is that we want to study the computational complexity of JA, and recognising a tautology or a contradiction is itself a computationally intractable problem. We write $\Phi^+$ for the set of non-negated formulas in $\Phi$, i.e., $\Phi = \Phi^+ \cup \{\neg\varphi \mid \varphi \in \Phi^+\}$.

**Definition 2.** *A **judgment set** $J$ for the agenda $\Phi$ is a subset $J \subseteq \Phi$.*

We call a judgment set $J$ *complete* if $\alpha \in J$ or $\sim\alpha \in J$ for all $\alpha \in \Phi$; we call it *complement-free*[1] if for all $\alpha \in \Phi$ it is not the case that both $\alpha$ and its complement are in $J$; and we call it *consistent* if there exists an assignment that makes all formulas in $J$ true. Let $\mathcal{J}(\Phi)$ denote the set of all complete and consistent subsets of $\Phi$.

We shall occasionally interpret a judgment set $J$ as a (characteristic) function $J : \Phi \rightarrow \{0, 1\}$ with $J(\varphi) = 1$ if $\varphi \in J$ and $J(\varphi) = 0$ if $\varphi \notin J$. The *Hamming distance* $H(J, J')$ between two (complete and complement-free) judgment sets $J$ and $J'$ is the number of positive formulas on which they differ:

$$H(J, J') \quad = \quad \sum_{\varphi \in \Phi^+} |J(\varphi) - J'(\varphi)|$$

Given a set $\mathcal{N} = \{1, \ldots, n\}$ of $n > 1$ *individuals* (or *agents*), we write $\boldsymbol{J} = (J_1, \ldots, J_n) \in \mathcal{J}(\Phi)^n$ for a generic *profile* of judgment sets, one for each individual.[2] For ease of exposition, we shall assume that $n$ is odd (i.e., $n \geqslant 3$). We write $N_\varphi^{\boldsymbol{J}} = \{i \in \mathcal{N} \mid \varphi \in J_i\}$ for the set of individuals accepting the formula $\varphi$ under profile $\boldsymbol{J}$.

**Definition 3.** *A (resolute) **judgment aggregation procedure** for the agenda $\Phi$ and the set of individuals $\mathcal{N}$ with $n = |\mathcal{N}|$ is a function $F : \mathcal{J}(\Phi)^n \rightarrow 2^\Phi$.*

---

1. This property is called *weak consistency* by Dietrich (2007), and *consistency* by List and Pettit (2002). Our choice of terminology is intended to emphasise the fact that it is a purely syntactic notion, not involving any model-theoretic concept, a distinction we believe is worth stressing.

2. In previous work we have used the more general notation $\mathcal{J}(\Phi)^{\mathcal{N}}$ (i.e., the set of functions from $\mathcal{N}$ to $\mathcal{J}(\Phi)$) for the set of admissible profiles (Endriss et al., 2010a). This is useful when $\mathcal{N}$ might be infinite or when we do not necessarily want to associate the set of individuals with a set of natural numbers, but we do not require this level of generality here.

That is, an aggregation procedure maps any profile of individual (complete and consistent) judgment sets to a single collective judgment set (an element of the powerset of $\Phi$). We shall occasionally refer to the assumption of all individual judgment sets being complete and consistent as *individual rationality*. Note that the collective judgment set need not be complete and consistent (that is, "collective rationality" need not hold). The kind of procedure defined above is called *resolute*, because it will return a single judgment set for any profile. Later, we shall also discuss *irresolute* JA procedures, which may return a nonempty set of judgment sets (that are tied for "winning"). Finally, note that, since $F$ is defined on the set of *all* profiles of consistent and complete judgment sets, we are implicitly making a *universal-domain* assumption, which is sometimes stated as a separate property (List & Pettit, 2002).

## 2.2 Axiomatic Properties

In Definition 3 we did not put any constraints on the collective judgment set, the outcome of the aggregation process. This is the role of the following definition:

**Definition 4.** *A judgment aggregation procedure $F$, defined on an agenda $\Phi$, is said to be:*
  *(i)* **complete** *if $F(\boldsymbol{J})$ is complete for every $\boldsymbol{J} \in \mathcal{J}(\Phi)^n$;*
 *(ii)* **complement-free** *if $F(\boldsymbol{J})$ is complement-free for every $\boldsymbol{J} \in \mathcal{J}(\Phi)^n$;*
*(iii)* **consistent** *if $F(\boldsymbol{J})$ is consistent for every $\boldsymbol{J} \in \mathcal{J}(\Phi)^n$.*

We now present several *axioms* to provide a normative framework in which to state what the desirable properties of an acceptable aggregation procedure should be. Note that not every procedure has to satisfy every axiom. Rather, axioms model desiderata that some procedures satisfy and others do not. The first axiom is a very basic requirement, restricting possible aggregators $F$ in terms of fundamental properties of the outcomes they produce.

**Weak Rationality** (WR): $F$ is complete and complement-free.[3]

This condition differs from what has been called "collective rationality" in the literature on JA (List & Puppe, 2009), as we do not require the collective judgment set to be consistent. The first reason not to include consistency in our most basic notion of rationality is that the requirements of (WR) are purely syntactic notions that can easily be checked automatically, which is not the case for consistency. The second reason is that consistency is not intrinsic to the aggregation function, but rather depends on the properties of the agenda. This point will be made more precise in Section 5, where we will study the consistency of a class of aggregators depending on the agenda.

  The following are the most important axioms discussed in the literature on JA (List & Pettit, 2002; Dietrich, 2007; List & Puppe, 2009; Nehring & Puppe, 2010):

**Unanimity** (U): If $\varphi \in J_i$ for all agents $i \in \mathcal{N}$, then $\varphi \in F(\boldsymbol{J})$.

---

3. In our previous work (Endriss et al., 2010a), we used a definition of weak rationality that in addition to completeness and complement-freeness also included the (very weak) technical requirement that no contradictory formula should be universally accepted under all profiles. As a consequence, some of our results later on are stated slightly differently.

**Anonymity** (A): For any profile $\boldsymbol{J}$ in $\mathcal{J}(\Phi)^n$ and any permutation $\sigma : \mathcal{N} \to \mathcal{N}$, we have $F(J_1, \ldots, J_n) = F(J_{\sigma(1)}, \ldots, J_{\sigma(n)})$.

**Neutrality** (N): For any two formulas $\varphi$, $\psi$ in the agenda $\Phi$ and any profile $\boldsymbol{J}$ in $\mathcal{J}(\Phi)^n$, if for all agents $i \in \mathcal{N}$ we have that $\varphi \in J_i \Leftrightarrow \psi \in J_i$, then $\varphi \in F(\boldsymbol{J}) \Leftrightarrow \psi \in F(\boldsymbol{J})$.

**Independence** (I): For any formula $\varphi$ in the agenda $\Phi$ and any two profiles $\boldsymbol{J}$, $\boldsymbol{J}'$ in $\mathcal{J}(\Phi)^n$, if $\varphi \in J_i \Leftrightarrow \varphi \in J_i'$ for all agents $i \in \mathcal{N}$, then $\varphi \in F(\boldsymbol{J}) \Leftrightarrow \varphi \in F(\boldsymbol{J}')$.

**Systematicity** (S): For any two formulas $\varphi$, $\psi$ in the agenda $\Phi$ and any two profiles $\boldsymbol{J}$, $\boldsymbol{J}'$ in $\mathcal{J}(\Phi)^n$, if $\varphi \in J_i \Leftrightarrow \psi \in J_i'$ for all agents $i \in \mathcal{N}$, then $\varphi \in F(\boldsymbol{J}) \Leftrightarrow \psi \in F(\boldsymbol{J}')$.

Unanimity expresses the idea that if all individuals accept a given judgment, then so should the collective.[4] Anonymity states that aggregation should be symmetric with respect to individuals, i.e., all individuals should be treated the same. Neutrality is a symmetry requirement for propositions: if the same subgroup accepts two propositions, then either both or neither should be collectively accepted. Independence says that if a proposition is accepted by the same subgroup under two otherwise distinct profiles, then that proposition should be accepted either under both or under neither profile. Systematicity is satisfied if and only if both neutrality and independence are. While all of these axioms are intuitively appealing, they are stronger than they may seem at first, and several impossibility theorems, establishing inconsistencies between certain combinations of axioms with other desiderata, have been proved in the literature. The original impossibility theorem of List and Pettit (2002), for instance, shows that (under certain assumptions regarding the agenda) there can be no complete and consistent aggregation procedure satisfying (A) and (S).

A further important property is monotonicity. We introduce two different axioms for monotonicity. The first is the one commonly used in the literature (Dietrich & List, 2007a; List & Puppe, 2009). It implicitly relies on the independence axiom. The second, introduced in our previous work (Endriss et al., 2010a), is designed to be applied to neutral procedures. For systematic procedures the two formulations are equivalent.

**I-Monotonicity** ($M^I$): For any formula $\varphi$ in the agenda $\Phi$ and any two profiles $\boldsymbol{J}$, $\boldsymbol{J}'$ in $\mathcal{J}(\Phi)^n$, if $\varphi \in J_i \Rightarrow \varphi \in J_i'$ for all agents $i \in \mathcal{N}$, and for some $s \in \mathcal{N}$ we have that $\varphi \notin J_s$ and $\varphi \in J_s'$, then $\varphi \in F(\boldsymbol{J}) \Rightarrow \varphi \in F(\boldsymbol{J}')$.

**N-Monotonicity** ($M^N$): For any two formulas $\varphi, \psi$ in the agenda $\Phi$ and any profile $\boldsymbol{J}$ in $\mathcal{J}(\Phi)^n$, if $\varphi \in J_i \Rightarrow \psi \in J_i$ for all agents $i \in \mathcal{N}$ and $\varphi \notin J_s$ and $\psi \in J_s$ for some $s \in \mathcal{N}$, then $\varphi \in F(\boldsymbol{J}) \Rightarrow \psi \in F(\boldsymbol{J})$.

That is, ($M^I$) expresses that if $\varphi$ is collectively accepted (in $\boldsymbol{J}$) and receives additional support (in $\boldsymbol{J}'$, from agent $s$), then it should continue to be collectively accepted. Axiom ($M^N$) says that if $\varphi$ is collectively accepted and $\psi$ is accepted by a strict superset of the individuals accepting $\varphi$, then $\psi$ should also be collectively accepted.

Axioms can be used to define different *classes* of aggregation procedures: Given an agenda $\Phi$ and a list of desirable properties AX provided in the form of axioms, we define $\mathcal{F}_\Phi[\text{AX}]$ to be the set of all procedures $F : \mathcal{J}(\Phi)^n \to 2^\Phi$ that satisfy the axioms in AX.

---

4. This notion of unanimity is stronger than another common formulation only requiring $\boldsymbol{J} = (J, \ldots, J)$ to imply $F(\boldsymbol{J}) = J$ (List & Puppe, 2009), but the two are equivalent under the assumption of (I).

### 2.3 Judgment Aggregation Procedures

Next, we define three concrete types of aggregation procedures.

#### 2.3.1 Uniform Quota Rules and the Majority Rule

An aggregation procedure $F$ for $n = |\mathcal{N}|$ individuals is a *quota rule* if for every formula $\varphi$ there exists a quota $q_\varphi \in \{0, \ldots, n+1\}$ such that $\varphi \in F(\boldsymbol{J})$ if and only if $|N_\varphi^{\boldsymbol{J}}| \geqslant q_\varphi$. The class of quota rules has been studied in depth by Dietrich and List (2007a). In this paper, we are interested in a particular class of quota rules:

**Definition 5.** *Given some $m \in \{0, \ldots, n+1\}$ and an agenda $\Phi$, the **uniform quota rule** with quota $m$ is the aggregation procedure $F_m$ with $\varphi \in F_m(\boldsymbol{J}) \Leftrightarrow |N_\varphi^{\boldsymbol{J}}| \geqslant m$.*

An aggregation procedure satisfies (A), (I), ($M^I$), and (N) if and only if it is a uniform quota rule; this fact follows immediately from a result by Dietrich and List (2007a), who use a slightly more narrow definition of quota rule. Provided $m \neq n + 1$, the uniform quota rule $F_m$ also satisfies (U).

A quota rule of special interest is the *majority rule.* The majority rule is the uniform quota rule with $m = \frac{n+1}{2}$; it accepts a formula whenever there are more individuals accepting it than there are rejecting it (recall that we did assume $n$ to be odd). Clearly, the majority rule is the only uniform quota rule that satisfies (WR).

#### 2.3.2 The Premise-Based Procedure

As we have seen in the introduction, the majority rule may fail to produce a consistent outcome. Two basic aggregation procedures that can be set up in a way so as to avoid this problem have been discussed in the JA literature from the very beginning, namely the *premise-based* and the *conclusion-based* procedure (Kornhauser & Sager, 1993; Dietrich & Mongin, 2010). The basic idea is to divide the agenda into premises and conclusions. Under the premise-based procedure, we apply the majority rule to the premises and then infer which conclusions to accept given the collective judgments regarding the premises;[5] under the conclusion-based procedure we directly ask the agents for their judgments on the conclusions and leave the premises unspecified in the collective judgment set. That is, the conclusion-based procedure does not result in complete outcomes (indeed, strictly speaking, it does not conform to Definition 3), and we shall not consider it here. The premise-based procedure, on the other hand, can be set up in a way that guarantees consistent and complete outcomes, which provides a usable procedure of some practical interest.

For many JA problems, it may be natural to divide the agenda into premises and conclusions. Let $\Phi = \Phi^p \uplus \Phi^c$ be an agenda divided into a set of premises $\Phi^p$ and a set of conclusions $\Phi^c$, each of which is closed under complementation.

---

5. This is what is commonly understood by "premise-based procedure". Dietrich and Mongin (2010), who call this rule *premise-based majority voting*, have also investigated a more general class of premise-based procedures in which the procedure used to decide upon the premises need not be the majority rule.

**Definition 6.** *The **premise-based procedure** PBP for $\Phi^p$ and $\Phi^c$ is the function mapping each profile $\boldsymbol{J} = (J_1, \ldots, J_n) \in \mathcal{J}(\Phi)^n$ to the following judgment set:*

$$\mathrm{PBP}(\boldsymbol{J}) \;=\; \Delta \cup \{\varphi \in \Phi^c \mid \Delta \models \varphi\},$$

$$\text{where } \Delta = \{\varphi \in \Phi^p \mid |N_\varphi^{\boldsymbol{J}}| \geqslant \frac{n+1}{2}\}$$

That is, $\Delta$ is the set of premises accepted by a (strict) majority; and the PBP will return this set $\Delta$ together with those conclusions $\varphi$ that logically follow from $\Delta$ ($\Delta \models \varphi$).

If we want to ensure that the PBP always returns judgment sets that are consistent and complete, then we have to impose certain restrictions:

- If we want to guarantee *consistency*, then we have to impose restrictions on the premises. It is well-known that the majority rule is guaranteed to be consistent if and only if the agenda $\Phi$ satisfies the so-called *median property*, i.e., if every inconsistent subset of $\Phi$ has itself an inconsistent subset of size $\leqslant 2$ (Nehring & Puppe, 2007; List & Puppe, 2009).[6] This result immediately transfers to the PBP: it is consistent if and only if the set of premises satisfies the median property.

- If we want to guarantee *completeness*, then we have to impose restrictions on the conclusions: for any assignment of truth values to the premises, the truth value of each conclusion has to be fully determined.

We shall see in Section 5 that deciding whether a set of formulas satisfies the median property is highly intractable. That is, in its most general form, deciding whether the PBP is a consistent aggregation procedure for a given agenda is a complex problem. For a meaningful analysis, we therefore make two additional assumptions. First, we assume that the agenda $\Phi$ is closed under propositional variables: $p \in \Phi$ for any propositional variable $p$ occurring within any of the formulas in $\Phi$. Second, we equate the set of premises with the set of literals. Clearly, the above-mentioned conditions for consistency and completeness are satisfied under these assumptions.

So, to summarise, the instance of the PBP we shall work with in this paper is defined as follows: Under the assumption that the agenda is closed under propositional variables, the PBP accepts a literal $\ell$ if and only if more individuals accept $\ell$ than do accept $\sim\ell$; and the PBP accepts a compound formula if and only if it is entailed by the accepted literals. For consistent and complete input profiles, and assuming that $n$ is odd, this leads to a resolute JA procedure that is consistent and complete. On the downside, the PBP violates most of the standard axioms typically considered, such as (N) and (I). It even violates (U):

|          | $p$ | $q$ | $r$ | $p \vee q \vee r$ |
|----------|-----|-----|-----|-------------------|
| Agent 1  | Yes | No  | No  | Yes               |
| Agent 2  | No  | Yes | No  | Yes               |
| Agent 3  | No  | No  | Yes | Yes               |
| PBP      | No  | No  | No  | No                |

In this example, all three agents unanimously accept $p \vee q \vee r$, but when we aggregate using the PBP, then we end up rejecting $p \vee q \vee r$, because each of the three premises is rejected.

---

6. We shall discuss this result in detail in Section 5.

### 2.3.3 The Distance-Based Procedure

The basic idea of a distance-based approach to aggregation is to select an outcome that, in some sense, minimises the distance to the input profile. This idea has been used extensively in both preference aggregation (Kemeny, 1959) and belief merging (Konieczny & Pino Pérez, 2002). The first example of a JA procedure based on a notion of distance was introduced by Pigozzi (2006), albeit under the restrictive assumption that the agenda is closed under propositional variables and that each compound formula will either be unanimously accepted or unanimously rejected by all agents. Most importantly, in Pigozzi's approach the syntactic information contained in the agenda was discarded by moving the aggregation from the level of formulas to the level of models. A syntactic variant of this procedure has later been defined by Miller and Osherson (2009), which these authors call the *Prototype-Hamming* rule. This is the distance-based procedure we shall define and analyse here. It is an *irresolute* procedure, returning a (nonempty) set of collective judgment sets.

**Definition 7.** *Given an agenda* $\Phi$, *the **distance-based procedure** DBP is the function mapping each profile* $\boldsymbol{J} = (J_1, \ldots, J_n) \in \mathcal{J}(\Phi)^n$ *to the following set of judgment sets:*

$$\mathrm{DBP}(\boldsymbol{J}) \quad = \quad \operatorname*{argmin}_{J \in \mathcal{J}(\Phi)} \sum_{i \in \mathcal{N}} H(J, J_i)$$

A collective judgment set under the DBP minimises the amount of disagreement with the individual judgment sets (i.e., it minimises the sum of the Hamming distances with all individual judgment sets). Note that in cases where the majority rule leads to a consistent outcome, the outcome of the DBP coincides with that of the majority rule (making it a resolute procedure over these profiles). We can combine the DBP with a *tie-breaking rule* to obtain a resolute procedure.

The DBP is complete and consistent by design: only judgment sets in $\mathcal{J}(\Phi)$ are considered candidates when searching for a solution. However, it violates most of the standard axiomatic properties when those are adapted to the case of irresolute JA procedures (Lang, Pigozzi, Slavkovik, & van der Torre, 2011). In particular, the DBP is not independent; indeed, it is based on the very idea that correlations between propositions should be exploited rather than neglected.

### 2.4 Representation Results

We now prove a number of representation results that characterise the aggregation procedures that satisfy certain combinations of axioms. All of the results in this section are known results, but—despite being very useful—they have rarely been stated explicitly in the literature.

Observe that an aggregation procedure $F$ satisfies (I) if and only if there exists a family of sets of *winning coalitions* $\mathcal{W}_\varphi \subseteq 2^{\mathcal{N}}$, one for each formula $\varphi \in \Phi$, such that $\varphi \in F(\boldsymbol{J}) \Leftrightarrow N_\varphi^{\boldsymbol{J}} \in \mathcal{W}_\varphi$. Imposing additional axioms, on top of (I), forces some additional structure onto the family of winning coalitions:

- $F$ satisfies (I) and (U) if and only if the grand coalition belongs to every set of winning coalitions: $\mathcal{N} \in \mathcal{W}_\varphi$.

- $F$ satisfies (I) and (N), i.e., it satisfies (S), if and only if there exists a *single* set of winning coalitions $\mathcal{W} \subseteq 2^{\mathcal{N}}$ such that $\varphi \in F(\boldsymbol{J}) \Leftrightarrow N_\varphi^{\boldsymbol{J}} \in \mathcal{W}$.

- $F$ satisfies (I) and (A) if and only if collective acceptance of a formula only depends on the *number* of individuals accepting it: $C \in \mathcal{W}_\varphi$ and $|C| = |C'|$ imply $C' \in \mathcal{W}_\varphi$.

One consequence of the latter two insights is that, if $F$ satisfies (A) and (S), then $|N_\varphi^{\boldsymbol{J}}| = |N_\psi^{\boldsymbol{J}'}|$ implies $\varphi \in F(\boldsymbol{J}) \Leftrightarrow \psi \in F(\boldsymbol{J}')$. This is a well-known fact; List and Pettit (2002), for instance, use it in the proof of their impossibility theorem (for the special case of $\boldsymbol{J} = \boldsymbol{J}'$). Note that a (somewhat surprising) consequence of this fact is that, in case $n$ is even, there exists no aggregation procedure that satisfies (A), (S), as well as (WR). To see this, it suffices to consider a (single) profile $\boldsymbol{J}$ where exactly $\frac{n}{2}$ agents accept $\varphi$ and $\frac{n}{2}$ agents accept $\neg\varphi$. Then $|N_\varphi^{\boldsymbol{J}}| = |N_{\neg\varphi}^{\boldsymbol{J}}|$, i.e., either both $\varphi$ and $\neg\varphi$ must be in $F(\boldsymbol{J})$, contradicting complement-freeness, or neither $\varphi$ nor $\neg\varphi$ must be in $F(\boldsymbol{J})$, this time contradicting completeness. We emphasise that this basic impossibility result does not involve any notion of logical consistency.

On the other hand, when $n$ is odd (which we shall continue to assume), then these axioms characterise a relevant class of aggregation procedures:

**Proposition 1.** $F \in \mathcal{F}_\Phi[\text{WR}, \text{A}, \text{S}]$ *if and only if there exists a function* $h : \{0, \ldots, n\} \to \{0, 1\}$ *satisfying* $h(i) = 1 - h(n-i)$ *for all* $i \in \mathcal{N}$ *such that* $\varphi \in F(\boldsymbol{J}) \Leftrightarrow h(|N_\varphi^{\boldsymbol{J}}|) = 1$.

*Proof.* We have already seen that when $F$ satisfies (S) and (A), then $|N_\varphi^{\boldsymbol{J}}| = |N_\psi^{\boldsymbol{J}'}|$ implies $\varphi \in F(\varphi) \Leftrightarrow \psi \in F(\boldsymbol{J}')$. The latter is equivalent to the existence of a function $h : \{0, \ldots, n\} \to \{0, 1\}$ with $\varphi \in F(\boldsymbol{J}) \Leftrightarrow h(|N_\varphi^{\boldsymbol{J}}|) = 1$. The additional requirement of $h(i) = 1 - h(n-i)$ then is a consequence of (WR). The other direction is immediate: as acceptance of a formula under $F$ only depends on the number of agents accepting it, $F$ must be anonymous, neutral and independent; the condition $h(i) = 1 - h(n-i)$ furthermore ensures completeness and complement-freeness. $\square$

Dropping either neutrality or independence, we obtain the following representation results:

**Proposition 2.** $F \in \mathcal{F}_\Phi[\text{WR}, \text{A}, \text{I}]$ *if and only if there exists a function* $h_\varphi : \{0, \ldots, n\} \to \{0, 1\}$ *for every formula* $\varphi \in \Phi$ *satisfying* $h_\varphi(i) = 1 - h_{\sim\varphi}(n-i)$ *for all* $i \in \mathcal{N}$ *such that* $\varphi \in F(\boldsymbol{J}) \Leftrightarrow h_\varphi(|N_\varphi^{\boldsymbol{J}}|) = 1$.

*Proof.* As is clear from our characterisation of procedures satisfying (I) and (A) in terms of winning coalitions given above, for such a procedure we can always decide whether $\varphi$ should be collectively accepted by only looking at the cardinality of the coalition accepting $\varphi$. The rest of the proof proceeds just as for Proposition 1. $\square$

**Proposition 3.** $F \in \mathcal{F}_\Phi[\text{WR}, \text{A}, \text{N}]$ *if and only if there exists a function* $h_{\boldsymbol{J}} : \{0, \ldots, n\} \to \{0, 1\}$ *for every profile* $\boldsymbol{J} \in \mathcal{J}(\Phi)^n$ *satisfying* $h_{\boldsymbol{J}}(i) = 1 - h_{\boldsymbol{J}}(n-i)$ *for all* $i \in \mathcal{N}$ *such that* $\varphi \in F(\boldsymbol{J}) \Leftrightarrow h_{\boldsymbol{J}}(|N_\varphi^{\boldsymbol{J}}|) = 1$.

*Proof.* When we drop (I), then winning coalitions are not anymore associated with formulas, but depend on the profile $\boldsymbol{J}$ we are in. (N) merely ensures that those winning coalitions do not also depend on the formula in question. (WR) again forces the symmetry requirement $h_{\boldsymbol{J}}(i) = 1 - h_{\boldsymbol{J}}(n-i)$. The opposite direction is once again immediate. $\square$

For each of the three representation results above, if we add (U) to the list of axioms, then this corresponds to requiring $h(n) = 1$ for each of the characteristic functions $h$.

Finally, recall that we have seen in Section 2.3.1, that $F \in \mathcal{F}_\Phi[\mathrm{A}, \mathrm{S}, \mathrm{M}^\mathrm{I}]$ if and only if $F$ is a uniform quota rule and that $F \in \mathcal{F}_\Phi[\mathrm{WR}, \mathrm{A}, \mathrm{S}, \mathrm{M}^\mathrm{I}]$ if and only if $F$ is the majority rule. That is, the representation results stated above all concern natural weakenings of the combination of axioms characterising the majority rule. In particular, we chose never to drop the anonymity axiom, because we find it very appealing and uncontroversial for JA. We also consider unanimity and weak rationality very fundamental (although we make exceptions for the class of quota rules). The independence and neutrality axioms, on the other hand, are much more debatable, which is why we have considered the various options of either including and not including them (although we always keep at least one of them, to maintain a minimal amount of structure). That is, the classes of aggregation procedures covered by the representation results above are all very natural to focus on.

## 3. Winner Determination

In this section we define the problem of winner determination of a given JA procedure as a decision problem, and we study the computational complexity of this problem for each of the procedures presented in Section 2.3.

### 3.1 Problem Definition

The problem of winner determination in voting theory is that of computing the election winner given a profile of preferences supplied by the voters. The corresponding decision problem asks, given a preference profile and a candidate, whether the given candidate is the winner of the election. In JA, we want to compute $F(\boldsymbol{J})$ for a given profile $\boldsymbol{J}$. For a resolute aggregation procedure $F$, we can formulate a corresponding decision problem by asking, for a given formula, whether it belongs to $F(\boldsymbol{J})$:

> WINDET($F$)
> **Instance:**   Agenda $\Phi$, profile $\boldsymbol{J} \in \mathcal{J}(\Phi)^n$, formula $\varphi \in \Phi$.
> **Question:**   Is $\varphi$ an element of $F(\boldsymbol{J})$?

By solving WINDET once for each formula in the agenda, we can compute the collective judgment set from an input profile. Note that asking instead whether a given judgment set $J^\star$ is equal to $F(\boldsymbol{J})$ does not lead to an appropriate formulation of the winner determination problem, because to actually compute the winner we would then have to solve our decision problem an exponential number of times (once for each possible $J^\star$).

For the case of irresolute JA procedures $F$ we can adapt the winner determination problem in the following way:

> WINDET$^\star$($F$)
> **Instance:**   Agenda $\Phi$, profile $\boldsymbol{J} \in \mathcal{J}(\Phi)^n$, subset $L \subseteq \Phi$.
> **Question:**   Is there a $J^\star \subseteq \Phi$ with $L \subseteq J^\star$ such that $J^\star \in F(\boldsymbol{J})$?

To see that this is an appropriate formulation of a decision problem corresponding to the task of computing *some* winning set, note that we can compute a winner using a polynomial number of queries to WinDet$^\star$ as follows. First, ask whether there exists a winning set including an arbitrarily chosen first formula of the agenda $\varphi_1$, i.e., $L = \{\varphi_1\}$. In case the answer is positive, consider a second formula $\varphi_2$ and query WinDet$^\star$ with $L = \{\varphi_1, \varphi_2\}$. Use subset $L = \{\sim\varphi_1, \varphi_2\}$ in case of a negative answer. Continue this process until all formulas in the agenda have been covered.[7]

## 3.2 Winner Determination for Quota Rules and the Premise-Based Procedure

It is immediately clear that winner determination is a polynomial problem for any quota rule, including the majority rule.

**Fact 4.** WinDet$(F_m)$ *is in* P *for any uniform quota rule* $F_m$.

Winner determination is also tractable for the premise-based procedure:

**Proposition 5.** WinDet(PBP) *is in* P.

*Proof.* Counting the number of agents accepting each of the premises and checking for each premise whether the positive or the negative instance has the majority is easy. This determines the collective judgment set as far as the premises are concerned. Deciding whether a given conclusion should be accepted by the collective now amounts to a model checking problem (is the conclusion $\varphi$ true in the model induced by the accepted premises/literals?), which can also be done in polynomial time. $\square$

## 3.3 Winner Determination for the Distance-Based Procedure

We now want to analyse the complexity of the winner determination problem for the distance-based procedure. As the DBP is irresolute, we study the decision problem WinDet$^\star$. As we shall see, WinDet$^\star$(DBP) is $\Theta_2^p$-complete, thus showing that this rule is very hard to compute. The class $\Theta_2^p$ (also known as $\Delta_2^p(O(\log n))$, P$^{\text{NP[log]}}$ or P$_{||}^{\text{NP}}$) is the class of problems that can be solved in polynomial time asking a logarithmic number of queries to an NP oracle or, equivalently, that can be solved in polynomial time asking a polynomial number of such queries in parallel (Wagner, 1987; Hemachandra, 1989). To obtain our result, we first have to devise an NP oracle that will then be used in the proof of $\Theta_2^p$-membership. We shall use the following problem:

> WinDet$^\star_K$(DBP)
> **Instance:** Agenda $\Phi$, profile $\boldsymbol{J} \in \mathcal{J}(\Phi)^n$, subset $L \subseteq \Phi$, $K \in \mathbb{N}$.
> **Question:** Is there a $J^\star \in \mathcal{J}(\Phi)$ with $L \subseteq J^\star$ such that $\sum_{i \in \mathcal{N}} H(J^\star, J_i) \leqslant K$?

That is, we ask whether there exists a judgment set $J^\star$ with a Hamming distance to the profile of at most $K$ that accepts all the formulas in $L$. In other words, rather than aiming at computing a winning judgment set, this problem merely allows us to compute a judgment set of a certain minimal quality (where quality is measured in terms of the Hamming distance). We now show that this problem lies in NP.[8]

---

7. In line with recent work by Hemaspaandra, Hemaspaandra, and Menton (2012), we can therefore argue that our formulation of the winner determination problem is the correct decision problem associated with the *search* problem of actually computing a winning judgment set.

**Lemma 6.** $\text{WinDet}^\star_K(\text{DBP})$ *is in* NP.

*Proof.* We show that $\text{WinDet}^\star_K(\text{DBP})$ can be modelled as an integer program (without objective function). This proves membership in NP (Papadimitriou, 1981). Suppose we want to answer an instance of $\text{WinDet}^\star_K(\text{DBP})$. The number of subformulas of propositions occurring in the agenda $\Phi$ is linear in the *size* (not *cardinality*) of $\Phi$. We introduce a binary decision variable for each of these subformulas: $x_i \in \{0, 1\}$ for the $i$th subformula.

We first write constraints that ensure that the chosen outcome will correspond to a consistent judgment set (i.e., that $J^\star \in \mathcal{J}(\Phi)$). Note that we can rewrite any formula in terms of negation, conjunction, and bi-implication without resulting in a superpolynomial (or even superlinear) increase in size. So we only need to show how to encode the constraints for these connectives. The following table indicates how to write these constraints:

| $\varphi_2 = \neg\varphi_1$ | $x_2 = 1 - x_1$ |
|---|---|
| $\varphi_3 = \varphi_1 \wedge \varphi_2$ | $x_3 \leqslant x_1$ and $x_3 \leqslant x_2$ and $x_1 + x_2 \leqslant x_3 + 1$ |
| $\varphi_3 = \varphi_1 \leftrightarrow \varphi_2$ | $x_1 + x_2 \leqslant x_3 + 1$ and $x_1 + x_3 \leqslant x_2 + 1$ |
| | and $x_2 + x_3 \leqslant x_1 + 1$ and $1 \leqslant x_1 + x_2 + x_3$ |

Before we continue, consider the following way of rewriting the sum of distances featuring in the definition of $\text{WinDet}^\star_K(\text{DBP})$:

$$
\begin{aligned}
\sum_{i \in \mathcal{N}} H(J^\star, J_i) &= \sum_{i=1}^{n} \sum_{\varphi \in \Phi^+} |J^\star(\varphi) - J_i(\varphi)| \\
&= \frac{1}{2} \cdot \sum_{\varphi \in \Phi} \sum_{i=1}^{n} |J^\star(\varphi) - J_i(\varphi)| \\
&= \frac{1}{2} \cdot \sum_{\varphi \in \Phi} |n \cdot J^\star(\varphi) - \sum_{i=1}^{n} J_i(\varphi)|
\end{aligned}
$$

We will need to bound this sum from above. Now suppose that variables $x_i$ with indices $i \in \{1, \ldots, m\}$ with $m = |\Phi|$ are those that correspond to the propositions that are elements of $\Phi$. Let $a_i = |N^{\boldsymbol{J}}_{\varphi_i}|$ be the number of individuals that accept the $i$th proposition in $\Phi$ (under $\boldsymbol{J}$). To compute a winner under the DBP, we need to find a consistent judgment set $J^\star$ (characterised by variables $x_1, \ldots, x_m$) that minimises the sum $|n \cdot x_1 - a_1| + \cdots + |n \cdot x_m - a_m|$. We do this by introducing an additional set of integer variables $y_i > 0$ for $i = 1, \ldots, m$. We can ensure that $y_i = |n \cdot x_i - a_i|$ by adding the following constraints:[9]

$$
\begin{aligned}
(\forall i \leqslant m) \quad n \cdot x_i - a_i &\leqslant y_i \\
(\forall i \leqslant m) \quad a_i - n \cdot x_i &\leqslant y_i
\end{aligned}
$$

---

8. Our proof not only establishes membership in NP, but also suggests how to implement a solver for this difficult problem. As pointed out by one anonymous reviewer, it is also possible to prove NP-membership more directly, using a certificate that consists of both $J^\star$ and a satisfying assignment for $J^\star$.

9. To be precise, these constraints only ensure $|n \cdot x_i - a_i| \leqslant y_i$. However, our next constraint will force the $y_i$ to be minimal.

Now the sum $\frac{1}{2} \cdot \sum_{i=1}^{m} y_i$ corresponds to the Hamming distance between the winning set and the profile. To ensure it does not exceed $K$, we can add the following constraint:

$$\frac{1}{2} \cdot \sum_{i=1}^{m} y_i \leqslant K$$

Finally, we need to ensure that all the formulas in the set $L \subseteq \Phi$ get accepted. We do this by adding one last set of constraints:

$$(\text{for all } i \text{ such that } \varphi_i \in L) \quad x_i = 1$$

Now, by construction, the integer program we have presented is feasible if and only if the instance of $\textsc{WinDet}_K^\star(\text{DBP})$ we have started out with should be answered in the positive. This completes the proof. □

To obtain an upper bound for the winner determination problem for the DBP, we can now use a standard construction. This first involves identifying the "best" value for $K$, and then deciding $\textsc{WinDet}_K^\star(\text{DBP})$ for that value of $K$. The latter can be done with a logarithmic number of queries to the problem the complexity of which we have analysed in Lemma 6. Together, this yields the desired upper bound:

**Lemma 7.** $\textsc{WinDet}^\star(\text{DBP})$ *is in* $\Theta_2^p$.

*Proof.* The problem $\textsc{WinDet}^\star(\text{DBP})$ asks whether there exists a *winning* judgment set that accepts all formulas in a given subset $L \subseteq \Phi$. Since the Hamming distance between a judgment set and a profile is bounded from above by a polynomial figure, we can solve this problem by performing a binary search over $K$ using a logarithmic number of queries to $\textsc{WinDet}_K^\star(\text{DBP})$.

More precisely, since $\sum_{i \in \mathcal{N}} H(J^\star, J_i)) \leqslant K^\star = \frac{|\Phi|}{2} \cdot |\mathcal{N}|$, a figure that is polynomial in the size of the problem description, we can ask a first query to $\textsc{WinDet}_K^\star(\text{DBP})$ with $K = \frac{K^\star}{2}$ and an empty subset of designated formulas. In case of a positive answer, we can continue the search with a new $K = \frac{K^\star}{4}$, otherwise we move to the higher half of the interval querying $\textsc{WinDet}_K^\star(\text{DBP})$ with $K = \frac{3}{4} \cdot K^\star$. This process ends after a logarithmic number of steps, providing the exact Hamming distance $K^w$ of a winning candidate from the profile $\boldsymbol{J}$ under consideration. It is now sufficient to run the problem $\textsc{WinDet}_K^\star(\text{DBP})$ with $K = K^w$ and subset $L$ as in the original instance of $\textsc{WinDet}^\star(\text{DBP})$ we wanted to solve. In case the answer is positive, since there cannot be a winning judgment set with Hamming distance strictly less than $K^w$, one of the winning judgment sets contains all formulas in $L$. On the other hand, in case of a negative answer all judgment sets containing $L$ have Hamming distance bigger than $K^w$, and thus cannot belong to the winning set. □

Next, we show that the upper bound established by Lemma 7 is tight. We exploit the similarity of the DBP to the Kemeny rule in preference aggregation to build on a known $\Theta_2^p$-hardness result by Hemaspaandra et al. (2005).

**Lemma 8.** $\textsc{WinDet}^\star(\text{DBP})$ *is* $\Theta_2^p$-*hard*.

*Proof.* We build a reduction from the problem Kemeny Winner, as defined in the work of Hemaspaandra et al. (2005).[10] An instance of this problem consists of a set of candidates $C$, a profile of linear preference orders $\boldsymbol{P} = (P_1, \ldots, P_n)$ over $C$, and a designated candidate $c \in C$. Define the *Kemeny score* of $c$ as the following expression:

$$KemenyScore(c, \boldsymbol{P}) = \min\{\textstyle\sum_{i=1}^{n} dist(P_i, Q) \mid Q \text{ is a linear order with } top(Q) = c\}$$

Here, $dist(P_i, Q)$ is the Hamming distance between two linear orders (defined as the number of ordered pairs of candidates on which they disagree) and $top(Q)$ is the most preferred candidate under preference order $Q$. Kemeny Winner asks whether the Kemeny score of $c$ is less than or equal to the Kemeny score of all other candidates $d \in C$.

We now build an instance of WinDet$^\star$(DBP) to decide this problem. Define an agenda $\Phi_C$ in the following way. First, add propositional variables $p_{ab}$ for all ordered pairs of distinct candidates $a, b$ in $C$; these variables can encode a linear order over $C$ as a binary relation (where $p_{ab}$ stands for $a \succ b$). Now we can describe the properties of a linear order by means of formulas of the form $p_{ab} \wedge p_{bc} \rightarrow p_{ac}$ and $p_{ab} \leftrightarrow \neg p_{ba}$. We include all of these formulas, for all $a, b, c \in C$, in $\Phi_C$. In fact, we include $m^2 + 1$ syntactic variants (where $m = |C|$) for each of them.[11] The figure $m^2 + 1$ is chosen to be higher than the maximal Hamming distance between any two linear orders (which is $m^2$).

Given a preference profile $\boldsymbol{P}$, we can build a judgment profile $\boldsymbol{J^P}$ by encoding each order $P_i$ over $C$ in a judgment set $J_i^{\boldsymbol{P}}$ over $\Phi_C$. For example, if agent 1's preference order is $a \succ b \succ c$, then $J_1^{\boldsymbol{P}}$ will include the set $\{p_{ab}, \neg p_{ba}, p_{bc}, \neg p_{cb}, p_{ac}, \neg p_{ca}\}$. In addition, each $J_i^{\boldsymbol{P}}$ will include all of the syntactic copies of all of the formulas encoding linear orders.

Observe that we have $dist(P_i, P_j) = H(J_i^{\boldsymbol{P}}, J_j^{\boldsymbol{P}})$ by construction. It is therefore sufficient to ask a query to WinDet$^\star$(DBP) using $\Phi_C$ as the agenda, $\boldsymbol{J^P}$ as the profile, and $L = \{p_{cd} \mid d \in C, c \neq d\}$ as the set of propositions to accept for sure, to obtain an answer to the initial Kemeny Winner instance with designated candidate $c$. If the winning ranking features $c$ as the top candidates (i.e., formulas $p_{cd}$ are accepted for all other candidates $d$), then its Kemeny score will be lower than or equal to that of all other candidates, providing a positive answer to the original problem. A key insight here is to notice that judgment sets encoding relations that are not linear orders will not be considered in the minimisation process, since every disagreement on one of the formulas encoding linear orders will cause a much greater loss in the Hamming distance than what can be gained by modifying the variables encoding the individual candidate rankings. □

Putting Lemma 7 and 8 together yields a complete characterisation of the complexity of winner determination under distance-based aggregation:

**Theorem 9.** WinDet$^\star$(DBP) *is* $\Theta_2^p$*-complete.*

Theorem 9 shows that the DBP is highly intractable. However, by adapting efficient heuristics developed for the Kemeny rule (which, as seen in the proof of Lemma 8, is closely related to the DBP) it may be possible to obtain an implementation of the DBP that achieves an acceptable performance in practice (Conitzer, Davenport, & Kalagnanam, 2006).

---

10. Hemaspaandra et al. (2005) work with preferences that are weak orders, but point out that their results remain valid when linear orders are used instead. To simplify presentation, we work with linear orders.
11. For instance, for the formula $\varphi$ we might use the syntactic variants $\varphi$, $\varphi \wedge \top$, $\varphi \wedge \top \wedge \top$, and so forth.

## 4. Strategic Manipulation

In the context of voting, an agent is said to be able to manipulate a voting rule when there exists a situation in which voting in a manner that does not truthfully reflect her preferences will result in an outcome that she prefers to the outcome that would be realised if she were to vote truthfully (Gaertner, 2006). What would constitute an appropriate definition of manipulation in the context of JA is not immediately clear, because in JA there is no notion of preference. However, by fixing a suitable notion of "closeness" on judgment sets, it is possible to build a preference ordering starting from an individual's initial judgment set. This is the approach followed by Dietrich and List (2007c) for JA and by Everaere et al. (2007) in the related setting of belief merging. It builds on the assumption that an agent's individual judgment set is also her most preferred outcome and amongst any two outcomes she will prefer the one that is "closer" to that most preferred outcome. We will measure "closeness" using the Hamming distance and we will call an aggregation procedure *F manipulable* if it permits a situation where an agent can change the outcome to get closer to her truthful judgment by reporting untruthfully.

Our main interest will be the computational complexity of deciding whether a given agent can successfully manipulate under a given profile. In this context, a result showing that manipulation is computationally intractable would count as a positive result. Specifically, we will study this problem for the premise-based procedure. We will not do so for the family of quota rules, because (as we shall see) it is impossible to manipulate a quota rule in the aforementioned sense. We will also not study the manipulation problem for the distance-based procedure, because (as we have seen) even the much more basic winner determination problem already is intractable for this procedure.

### 4.1 Problem Definition

We first need to define a preference ordering over judgment sets for each agent $i \in \mathcal{N}$. In principle, there are any number of ways of doing this, but one reasonable approach is to assume that agent $i$'s judgment set $J_i$ is also her most preferred outcome and that her preferences over other outcomes depend on how close they are to $J_i$ (Dietrich & List, 2007c). We shall measure closeness using the Hamming distance, but other distances could also be used to this end (Duddy & Piggins, 2012). So we will say that agent $i$ *prefers* $J$ to $J'$ if and only if $H(J_i, J) < H(J_i, J')$.

Below we employ standard game-theoretical notation and denote by $(\boldsymbol{J}_{-i}, J_i')$ the profile that is like $\boldsymbol{J}$, except that the judgment set of agent $i$ has been replaced by $J_i'$.

**Definition 8.** $F$ is **manipulable** at profile $\boldsymbol{J} \in \mathcal{J}(\Phi)^n$ by agent $i \in \mathcal{N}$, if there exists an alternative judgment set $J_i' \in \mathcal{J}(\Phi)$ such that $H(J_i, F(\boldsymbol{J}_{-i}, J_i')) < H(J_i, F(\boldsymbol{J}))$.

That is, by reporting $J_i'$ rather than her truthful judgment set $J_i$, agent $i$ can achieve the outcome $F(\boldsymbol{J}_{-i}, J_i')$ and that outcome is closer (in terms of the Hamming distance) to her truthful (and most preferred) set $J_i$ than the outcome $F(\boldsymbol{J})$ that would get realised if she were to truthfully report $J_i$. A procedure that is not manipulable at any profile by any agent is called *strategy-proof*.

Dietrich and List (2007c) have shown that a JA procedure is strategy-proof if and only if it satisfies (I) and (M$^\text{I}$). Indeed, this follows immediately from our definitions: independence means that the would-be manipulator can consider one proposition at a time; monotonicity then means that it is always in her best interest to drive up the support for formulas in her judgment set and to reduce the support for those not in her judgment set, i.e., it is in her best interest to report her judgment set truthfully.[12]

For aggregation procedures for which strategy-proofness cannot be guaranteed, we want to study the algorithmic problem of computing a manipulating judgment set. To this end, we formulate manipulation as a decision problem for an aggregation procedure $F$:

> Manip($F$)
> **Instance:** Agenda $\Phi$, profile $\boldsymbol{J} \in \mathcal{J}(\Phi)^n$, agent $i \in \mathcal{N}$.
> **Question:** Is there a $J_i' \in \mathcal{J}(\Phi)$ such that $H(J_i, F(\boldsymbol{J}_{-i}, J_i')) < H(J_i, F(\boldsymbol{J}))$?

Note that we are asking *whether* an agent can manipulate successfully, rather than *how*. That is, this problem does not immediately correspond to the practical (and potentially harder) problem of computing an actual strategy for the manipulator. However, since the interest here is in obtaining *intractability* results (to provide protection against manipulation), we can safely concentrate on this formulation, which provides a lower bound for the corresponding search problem.

As we have seen, the uniform quota rules (including the majority rule) are all independent and monotonic, which means that they are also strategy-proof (so the algorithmic problem of deciding Manip does not arise for these procedures). Of course, this comes at the price of not always producing outcomes that are consistent.

## 4.2 Strategic Manipulation under the Premise-Based Procedure

We now prove that manipulating the premise-based procedure is intractable, thus showing the existence of the kind of a "jump" in computational complexity between winner determination and manipulation that is desirable in this context.

**Theorem 10.** Manip(PBP) *is* NP-*complete.*

*Proof.* We first establish NP-membership. An untruthful judgment set $J_i'$ yielding a preferred outcome can serve as a certificate. Checking the validity of such a certificate means checking that (a) $J_i'$ is actually a complete and consistent judgment set and that (b) the outcome produced by $J_i'$ is better than the outcome produced by the truthful set $J_i$. As for (a), checking completeness is easy. Consistency can also be decided in polynomial time: for every propositional variable $p$ in the agenda, $J_i'$ must include either $p$ or $\neg p$; this admits only a single possible model; all that remains to be done is checking that all compound formulas in $J_i'$ are satisfied by that model.[13] As for (b), we need to compute the outcomes for $J_i$ and $J_i'$ (by Proposition 5, this is polynomial), compute their Hamming distances from $J_i$, and compare those two distances.

Next, we prove NP-hardness by reducing Sat to Manip(PBP). Suppose we are given a propositional formula $\varphi$ and want to check whether it is satisfiable. We will build a

---

12. Note that this does not contradict the Gibbard-Satterthwaite Theorem in voting theory (Gaertner, 2006). That theorem involves a universal-domain assumption, while the manner in which we are using the Hamming distance to induce preferences from judgment sets amounts to a domain restriction.

13. That is, at this point we crucially rely on our assumption that the PBP is only defined for agendas that are closed under propositional variables.

judgment profile for three agents such that the third agent can manipulate the aggregation if and only if $\varphi$ is satisfiable. Let $p_1, \ldots, p_m$ be the propositional variables occurring in $\varphi$, and let $q_1, q_2$ be two additional propositional variables. Define an agenda $\Phi$ that contains all atoms $p_1, \ldots, p_m, q_1, q_2$ and their negation, as well as $m + 2$ syntactic variants of the formula $q_1 \vee (\varphi \wedge q_2)$, as well as the complements of all of these formulas. For instance, if $\psi = q_1 \vee (\varphi \wedge q_2)$, we might use the syntactic variants $\psi$, $\psi \wedge \top$, $\psi \wedge \top \wedge \top$, and so forth. Now consider the profile $\boldsymbol{J}$ below (with the rightmost column having a "weight" of $m + 2$):

|         | $p_1$ | $p_2$ | $\cdots$ | $p_m$ | $q_1$ | $q_2$ | $q_1 \vee (\varphi \wedge q_2)$ |
|---------|-------|-------|----------|-------|-------|-------|---------------------------------|
| $J_1$   | 1     | 1     | $\cdots$ | 1     | 0     | 0     | ?                               |
| $J_2$   | 0     | 0     | $\cdots$ | 0     | 0     | 1     | ?                               |
| $J_3$   | 1     | 1     | $\cdots$ | 1     | 1     | 0     | 1                               |
| $F(\boldsymbol{J})$ | 1 | 1 | $\cdots$ | 1 | 0 | 0 | 0                               |

The judgments of agents 1 and 2 regarding $q_1 \vee (\varphi \wedge q_2)$ are irrelevant for our argument, so they are indicated as "?" in the table (but note that they can be determined in polynomial time; in particular, $J_1(q_1 \vee (\varphi \wedge q_2)) = 0$ for any $\varphi$).

If agent 3 reports her judgment set truthfully (as shown in the table), then the Hamming distance between $J_3$ and the collective judgment set will be $1 + (m + 2) = m + 3$. Note that agent 3 is decisive about all propositional variables (i.e., premises) except for $q_1$ (which will certainly get rejected). Now:

- If $\varphi$ is satisfiable, then agent 3 can report judgments regarding $p_1, \ldots, p_m$ that correspond to a satisfying assignment for $\varphi$. If she furthermore accepts $q_2$, then all $m + 2$ copies of $q_1 \vee (\varphi \wedge q_2)$ will get accepted in the collective judgment set. Thus, the Hamming distance from $J_3$ to this new outcome will be at most $m + 2$, i.e., agent 3 will have manipulated successfully.

- If $\varphi$ is not satisfiable, then there is no way to get any of the $m+2$ copies of $q_1 \vee (\varphi \wedge q_2)$ accepted (and $q_1$ will get rejected in any case). Thus, agent 3 has no means of improving over the Hamming distance of $m + 3$ she can guarantee for herself by reporting truthfully.

Hence, $\varphi$ is satisfiable if and only if agent 3 can manipulate successfully, and our reduction from Sat to Manip(PBP) is complete. □

Thus, manipulating the PBP is significantly harder than using it, at least in terms of worst-case complexity (and under the assumption that $P \neq NP$).

## 5. Safety of the Agenda

In this section, we introduce the concept of *safety of the agenda*: An agenda $\Phi$ is safe for a given aggregation procedure $F$, if the collective judgment set returned by $F$ will be consistent for any (consistent) input profile. Of course, this question is only relevant for aggregation procedures that are not always consistent to begin with, which is why we do not consider the PBP and the DBP in this section. In fact, our main interest will be in the safety of the agenda for entire classes of aggregation procedures, characterised by a set of axioms AX: $\Phi$ is safe for a class $\mathcal{F}_\Phi[\text{AX}]$ of aggregation procedures if it is safe for every procedure $F \in \mathcal{F}_\Phi[\text{AX}]$.

After defining the problem and relating it to so-called *agenda characterisation results* (or *possibility theorems*, as we shall call them) studied in the JA literature, we characterise safe agendas for a number of natural combinations of axioms and we establish the computational complexity of checking the safety of an agenda for these cases.

## 5.1 Problem Definition

When performing an aggregation of judgments, we would like to avoid paradoxical outcomes, i.e., we would like to ensure that the collective judgment set will be consistent. Whether or not this will indeed be the case depends on several factors: the aggregation procedure, the agenda, and the individual judgment sets. We cannot control what choices the individuals will make. We might not even know what aggregation procedure exactly they are going to use; we might only know about some of its properties, i.e., we might only know that it belongs to a certain class of procedures. Can we nevertheless guarantee that the collective judgment set will be consistent? We formalise this question as follows:

**Definition 9.** *An agenda $\Phi$ is **safe** with respect to a class of aggregation procedures $\mathcal{F}$, if every procedure in $\mathcal{F}$ is consistent when applied to profiles of judgment sets over $\Phi$.*

The example for a paradox presented in the introduction demonstrates the *un*safety of the agenda $\{p, \neg p, q, \neg q, p \wedge q, \neg(p \wedge q)\}$ with respect to the majority rule. The agenda $\{p, \neg p\}$, on the other hand, is immediately seen to be safe with respect to the full class of all weakly rational aggregation procedures.

The question of whether an agenda is safe is closely related to the rich literature on so-called *agenda characterisation results* (see, e.g., Nehring & Puppe, 2007; Dokow & Holzman, 2010; Dietrich & List, 2007b; List & Puppe, 2009). These authors have asked the following kind of question: for a given agenda and a given list of axiomatic requirements (always including the requirement of consistency), is it possible to find an aggregation procedure that meets those requirements on that agenda? We may rephrase this question as follows: given an agenda $\Phi$ and a list of axioms AX (now excluding consistency), is it possible to find a procedure in $\mathcal{F}_\Phi[\text{AX}]$ that is consistent? To distinguish results of this kind from our *safety theorems* below (which are also agenda characterisations of a kind), we shall refer to them as *possibility theorems*. To summarise: while a possibility theorem shows that there is *some* consistent procedure in $\mathcal{F}_\Phi[\text{AX}]$, a safety theorem shows that *all* procedures in $\mathcal{F}_\Phi[\text{AX}]$ are consistent.

Note that in case a "class" of aggregation procedures consists of just a single aggregation procedure (e.g., $\mathcal{F}_\Phi[\text{WR}, \text{A}, \text{S}, \text{M}^\text{I}]$ consists only of the majority rule), possibility and safety results coincide.

Possibility theorems are important from the point of view of the mechanism designer: given certain axioms that I would like to see satisfied, is it still possible to design an aggregation procedure meeting them once I know certain characteristics of the kind of agenda on which the procedure should be used? That is, this is a question we are likely to ask in an "off-line" situation and only once. Safety theorems, on the other hand, are more likely to play a role in an "on-line" situation and they arguably are of particular interest for applications. The reason is that actual users are more likely to want an assurance that aggregation *will* be consistent (provided certain axioms are satisfied and the agenda has certain properties) rather than to learn that there *exists* a consistent form of aggregation (satisfying certain axioms). For instance, suppose we want to give certain guarantees for the quality of operations of a multiagent system, but without full knowledge of the precise specification of every individual agent and without full knowledge of all the interaction protocols they are going to employ. We might nevertheless have sufficient information for a safety theorem to apply, in which case we can check, for a given agenda, whether consistency can be guaranteed. That is, deciding whether safety holds is a question we might have to answer again and again, for many different agendas. This is why the computational complexity of this problem is a relevant question.

## 5.2 Agenda Properties

As we shall see, if an agenda satisfies certain structural properties, then that might be a sufficient condition to ensure safety with respect to certain aggregation rules. It turns out that the types of agenda properties that are of help here are similar to those that feature in known possibility theorems. Specifically, we shall make use of the so-called *median property*, introduced by Nehring and Puppe (2007).[14]

**Definition 10.** *We say that an agenda* $\Phi$ *satisfies the **median property** (MP), if every inconsistent subset of* $\Phi$ *has itself an inconsistent subset of size at most 2.*

In other words, $\Phi$ satisfies the MP if it has no *minimally inconsistent subset* (*mi-subset*) with more than 2 elements. Note that in case $\Phi$ is known not to include any tautologies (and thus no contradictions), this definition simplifies to requiring that any mi-subset must be exactly of size 2. We can generalise the median property as follows:

**Definition 11.** *Let* $k \geqslant 2$. *An agenda* $\Phi$ *satisfies the **k-median property** (kMP), if every inconsistent subset of* $\Phi$ *has itself an inconsistent subset of size at most k.*

That is, the MP and the 2MP are the same property. Agendas satisfying the MP are already quite simple, but the restriction can be made tighter by requiring all inconsistent subsets to have a particular form. In the sequel, we call an inconsistent set $\Delta$ *nontrivially inconsistent* if it does not contain any single formula that is a contradiction.

**Definition 12.** *An agenda* $\Phi$ *satisfies the **simplified median property** (SMP), if every nontrivially inconsistent subset of* $\Phi$ *has a subset of the form* $\{\varphi, \psi\}$ *with* $\varphi$ *being logically equivalent to* $\neg\psi$.

A further simplification yields:

**Definition 13.** *An agenda* $\Phi$ *satisfies the **syntactic simplified median property** (SSMP), if every nontrivially inconsistent subset of* $\Phi$ *has a subset of the form* $\{\varphi, \neg\varphi\}$.

---

14. The name *median property* derives from the work of Nehring and Puppe (2007), who analyse social choice functions for a class of vector spaces called median spaces.

Agendas satisfying the SSMP are composed of uncorrelated formulas, i.e., they are essentially equivalent to agendas composed of atoms alone. The SMP is less restrictive, allowing for logically equivalent but syntactically different formulas.

Observe that every agenda that satisfies the SMP also satisfies the MP. The converse is not true: $\Phi = \{p, \neg p, p \wedge q, \neg(p \wedge q)\}$ satisfies the MP, but not the SMP. Similarly, every agenda that satisfies the SSMP also satisfies the SMP. Again, the converse is not true: $\Phi = \{p, \neg p, p \wedge p, \neg(p \wedge p)\}$ satisfies the SMP, but not the SSMP.

### 5.3 Safety Theorems: Linking Agenda Properties and Axioms

We now prove several characterisation results for the safe aggregation of judgments, concentrating on classes of procedures defined by weakening the axiomatisation of the majority rule. We begin with a safety theorem for the majority rule itself. In fact, this result is familiar from the literature (Nehring & Puppe, 2007), although it is presented there in a different form. Despite the fact that it is a known result, we still provide a proof, which arguably is simpler than translating the result of Nehring and Puppe into our setting.

**Theorem 11.** *An agenda $\Phi$ is safe for the majority rule if and only if $\Phi$ satisfies the* MP.

*Proof.* Let $F$ be the majority rule.

($\Leftarrow$) First, suppose $\Phi$ satisfies the MP. We need to show that $F(\boldsymbol{J})$ is consistent for any $\boldsymbol{J} \in \mathcal{J}(\Phi)^n$. For the sake of contradiction, suppose it is not, and let $\Delta$ be a mi-subset of $F(\boldsymbol{J})$. As $F(\boldsymbol{J}) \subseteq \Phi$, $\Delta$ can have at most 2 elements. Clearly, it cannot be the case that $F(\boldsymbol{J})$ includes a contradiction $\varphi^{\perp}$, as that would mean that a majority of the agents would have accepted $\varphi^{\perp}$. Hence, $\Delta$ must be a set of exactly two formulas, say, $\varphi$ and $\psi$. This means that $\varphi$ must have been accepted by $\frac{n+1}{2}$ or more agents and $\psi$ must have been accepted by $\frac{n+1}{2}$ or more agents. Hence, by the pigeon hole principle, at least one agent must have accepted both of them, thereby contradicting individual rationality.

($\Rightarrow$) For the other direction, suppose $\Phi$ does *not* satisfy the MP, i.e., $\Phi$ has a mi-subset $\Delta$ of size $k \geqslant 3$. We need to show that there exists a profile $\boldsymbol{J}$ such that $F(\boldsymbol{J})$ is inconsistent. Let $\varphi$ and $\psi$ be two distinct formulas in $\Delta$. Now consider a profile $\boldsymbol{J}$ with the following properties (recall that we assume that $n \geqslant 3$): (1) the first $\frac{n-1}{2}$ agents accept all formulas in $\Delta$ except for $\varphi$; (2) the last $\frac{n-1}{2}$ agents accept all formulas in $\Delta$ except for $\psi$; and (3) the "middlemost" agent $\frac{n+1}{2}$ accepts $\varphi$ and $\psi$ and no other formula in $\Delta$. That is, no individual agent accepts all of the formulas in $\Delta$, i.e., we really can build an individually rational profile with these properties (note that any consistent subset of $\Delta$ can always be extended to a complete and consistent judgment set in $\Phi$). However, under this profile each of the formulas in $\Delta$ has a majority and we get $\Delta \subseteq F(\boldsymbol{J})$, i.e., $F(\boldsymbol{J})$ is inconsistent. $\square$

The reason that in this case we were able to rely on a known result is the aforementioned fact that for classes of aggregation procedures consisting of just a single procedure, safety and possibility results coincide. Unfortunately, for larger classes of procedures, this approach of exploiting known possibility results cannot be used.

We first establish two *sufficient* conditions for the safety of the agenda, for two different (fairly large) classes of aggregation procedures:

**Lemma 12.** *If an agenda $\Phi$ satisfies the* SSMP, *then $\Phi$ is safe for $\mathcal{F}_{\Phi}[\mathrm{WR}, \mathrm{U}]$.*

*Proof.* Consider an aggregation procedure that satisfies (WR) and (U). Let $\Phi$ be an agenda that satisfies the SSMP. Hence, the only way to obtain an inconsistent outcome would be to either accept an inconsistent formula or to accept a formula $\varphi$ and its syntactic complement $\neg\varphi$. The latter possibility is excluded by (WR). So, for the sake of excluding also the former possibility, suppose the inconsistent formula $\varphi^{\perp}$ has been collectively accepted. By individual rationality, $\sim\varphi^{\perp}$ will get accepted by all agents. Hence, by (U), $\sim\varphi^{\perp}$ will be collectively accepted, and thus $\varphi^{\perp}$ will be collectively rejected by (WR). □

**Lemma 13.** *If an agenda $\Phi$ satisfies the* SMP*, then $\Phi$ is safe for $\mathcal{F}_{\Phi}[\mathrm{WR}, \mathrm{U}, \mathrm{N}]$.*

*Proof.* Let $F$ be an aggregation procedure that satisfies (WR), (U) and (N), and let $\Phi$ be an agenda that satisfies the SMP. For the sake of contradiction, suppose there exists a profile $\boldsymbol{J} \in \mathcal{J}(\Phi)^n$ such that $F(\boldsymbol{J})$ is inconsistent. We distinguish two cases:

(1) There exists a set $\{\varphi, \psi\} \subseteq F(\boldsymbol{J})$ with $\varphi$ being logically equivalent to $\sim\psi$. But given that all individual judgment sets are assumed to be complete and consistent, $\varphi$ being logically equivalent to $\sim\psi$ means that every agent who accepts $\varphi$ will also accept $\sim\psi$, and *vice versa*, i.e., $N_{\varphi}^{\boldsymbol{J}} = N_{\sim\psi}^{\boldsymbol{J}}$. Together with (N) this entails $\varphi \in F(\boldsymbol{J}) \Leftrightarrow \sim\psi \in F(\boldsymbol{J})$. We already know that $\varphi \in F(\boldsymbol{J})$; thus, we also get $\sim\psi \in F(\boldsymbol{J})$. But as we also have $\psi \in F(\boldsymbol{J})$, we have obtained a contradiction to (WR).

(2) There exists an inconsistent formula $\varphi^{\perp} \in F(\boldsymbol{J})$. By the same argument as used in the proof of Lemma 12, this contradicts our assumption of $F$ satisfying (U) and (WR).

That is, we obtain a contradiction in all possible cases. □

Next, we prove two results concerning *necessary* conditions for the safety of the agenda (now we aim for relatively narrowly defined classes of aggregation procedures):

**Lemma 14.** *If an agenda $\Phi$ is safe for $\mathcal{F}_{\Phi}[\mathrm{WR}, \mathrm{A}, \mathrm{U}, \mathrm{S}]$, then $\Phi$ satisfies the* SMP*.*

*Proof.* Let $\Phi$ be an agenda that violates the SMP. We need to provide an example for an aggregation procedure $F$ that satisfies (WR), (A), (U) and (S) that will produce an inconsistent outcome for at least one input profile. We distinguish two cases:

(1) Suppose $\Phi$ violates the SMP by virtue of having a mi-subset of size greater than 2. In this case $\Phi$ also violates the MP. Then Theorem 11 shows that $\Phi$ is not safe for the majority rule. As the majority rule satisfies (WR), (A), (U) and (S), we are done.

(2) The only other possibility is for $\Phi$ to have a mi-subset consisting of two formulas that are not logical complements, i.e., there exists a set of the form $\{\varphi, \psi\} \subseteq \Phi$ with $\varphi \models \sim\psi$ but $\sim\psi \not\models \varphi$.[15] Consider then the following weakly rational, anonymous, unanimous and systematic aggregation procedure $F_h$ for 3 individuals, defined using the notation of Proposition 1: $h(0) = h(2) = 0$ and $h(1) = h(3) = 1$. That is, $F_h$ accepts a proposition only if it is accepted by an odd number of individuals.[16] Consider the following profile, restricted to $\varphi$ and $\psi$ and their complements: $J_1 = \{\sim\varphi, \sim\psi\}$, $J_2 = \{\varphi, \sim\psi\}$, $J_3 = \{\sim\varphi, \psi\}$. Note that each of these sets is consistent. However, the profile (opportunely extended to a profile on the whole agenda) will generate an inconsistent outcome, since both $\varphi$ and $\psi$ are accepted by exactly one individual.

---

15. For example, $\varphi$ might be $p \wedge q$ and $\psi$ might be $\neg p$.
16. This *parity rule* has also been used by Dokow and Holzman (2010) to provide a witness for one of their possibility results.

Hence, in all cases $\Phi$ fails to be safe for at least one procedure in $\mathcal{F}_\Phi[\mathrm{WR}, \mathrm{A}, \mathrm{U}, \mathrm{S}]$. $\qquad\square$

**Lemma 15.** *If an agenda $\Phi$ is safe for $\mathcal{F}_\Phi[\mathrm{WR}, \mathrm{A}, \mathrm{U}, \mathrm{I}]$, then $\Phi$ satisfies the* SSMP.

*Proof.* Let $\Phi$ be an agenda that violates the SSMP. If it also violates the SMP, then Lemma 14 applies and we are done.

Otherwise, there must be two formulas $\varphi$ and $\psi$ in $\Phi$ such that $\models \varphi \leftrightarrow \sim\psi$ but $\varphi \neq \sim\psi$, i.e., they are logical but not syntactic complements. Let $F$ be the procedure that accepts $\varphi$ (and rejects $\sim\varphi$) if at least one agent accepts $\varphi$, that accepts $\psi$ (and rejects $\sim\psi$) if at least one agent accepts $\psi$, and that behaves like the majority rule with respect to all other propositions. $F$ satisfies (WR), (A), (U) and (I), but $\Phi$ is not safe for $F$, because in case one agent accepts $\varphi$ and another $\psi$, the collective judgment set will include both $\varphi$ and $\psi$. $\quad\square$

We are now ready to state and prove our safety theorems:

**Theorem 16.** *An agenda $\Phi$ is safe for $\mathcal{F}_\Phi[\mathrm{WR}, \mathrm{A}, \mathrm{U}, \mathrm{S}]$ if and only if $\Phi$ satisfies the* SMP.

*Proof.* One direction is given by Lemma 14. The other follows from Lemma 13 together with the observation that $\mathcal{F}_\Phi[\mathrm{WR}, \mathrm{U}, \mathrm{N}] \supset \mathcal{F}_\Phi[\mathrm{WR}, \mathrm{A}, \mathrm{U}, \mathrm{S}]$. $\qquad\square$

This characterisation of safe agendas remains intact when we widen the class of aggregation procedures under consideration from systematic to neutral procedures:

**Theorem 17.** *An agenda $\Phi$ is safe for $\mathcal{F}_\Phi[\mathrm{WR}, \mathrm{A}, \mathrm{U}, \mathrm{N}]$ if and only if $\Phi$ satisfies the* SMP.

*Proof.* One direction follows from Lemma 14 together with the fact that $\mathcal{F}_\Phi[\mathrm{WR}, \mathrm{A}, \mathrm{U}, \mathrm{S}] \subset \mathcal{F}_\Phi[\mathrm{WR}, \mathrm{A}, \mathrm{U}, \mathrm{N}]$; the other from Lemma 13 and $\mathcal{F}_\Phi[\mathrm{WR}, \mathrm{U}, \mathrm{N}] \supset \mathcal{F}_\Phi[\mathrm{WR}, \mathrm{A}, \mathrm{U}, \mathrm{N}]$. $\quad\square$

Indeed, while Theorems 16 and 17 state safety results for particularly natural classes of aggregation procedures, by the same argument we can easily see that for *any* class $\mathcal{F}$ with $\mathcal{F}_\Phi[\mathrm{WR}, \mathrm{A}, \mathrm{U}, \mathrm{S}] \subseteq \mathcal{F} \subseteq \mathcal{F}_\Phi[\mathrm{WR}, \mathrm{U}, \mathrm{N}]$ it is the case that $\Phi$ is safe for $\mathcal{F}$ if and only if $\Phi$ satisfies the SMP.

If we drop neutrality from $\mathcal{F}_\Phi[\mathrm{WR}, \mathrm{A}, \mathrm{U}, \mathrm{S}]$ rather than independence, then we obtain an even more restrictive characterisation of safe agendas:

**Theorem 18.** *An agenda $\Phi$ is safe for $\mathcal{F}_\Phi[\mathrm{WR}, \mathrm{A}, \mathrm{U}, \mathrm{I}]$ if and only if $\Phi$ satisfies the* SSMP.

*Proof.* One direction is given by Lemma 15; the other follows from Lemma 12 together with $\mathcal{F}_\Phi[\mathrm{WR}, \mathrm{U}] \supset \mathcal{F}_\Phi[\mathrm{WR}, \mathrm{A}, \mathrm{U}, \mathrm{I}]$. $\qquad\square$

Again, we can generalise the above result to say that, for any class $\mathcal{F}$ with $\mathcal{F}_\Phi[\mathrm{WR}, \mathrm{A}, \mathrm{U}, \mathrm{I}] \subseteq \mathcal{F} \subseteq \mathcal{F}_\Phi[\mathrm{WR}, \mathrm{U}]$, it is the case that $\Phi$ is safe for $\mathcal{F}$ if and only if $\Phi$ satisfies the SSMP.

Finally, for uniform quota rules a characterisation result of the kind we seek is available in the literature (albeit under a different name), at least for rules with certain bounds imposed on the quota (Dietrich & List, 2007a). We state this interesting result as follows (recall that $n$ is the number of individuals):

**Theorem 19.** *Let $k \geqslant 2$. An agenda $\Phi$ is safe for the class of uniform quota rules $F_m$ satisfying the constraint $m > n - \frac{n}{k}$ if and only if $\Phi$ satisfies the $k$MP.*

Theorem 19 is a reformulation of Corollary 2(a) in the work of Dietrich and List (2007a) and we shall not prove it here.

Let us conclude this presentation of safety theorems with a remark on the role of the axiom (U) in our results above. Recall that we have not made any assumption about the agenda not including any contradictory formulas (or their complements, i.e., tautologies). If we do make this assumption (which is very common in the JA literature and certainly not unreasonable), then we can remove all mentionings of (U) in the safety results above. Indeed, we only ever used (U) in our proofs to avoid situations where a contradiction gets unanimously rejected yet collectively accepted. If we do not wish to make any assumption regarding the absence of contradictory formulas from the agenda, then we can still remove all mentionings of (U) from our safety results above, provided we replace all mentionings of the SMP with the property of both satisfying the SMP and not including any contradictory formulas (and accordingly for results involving the SSMP).

### 5.4 Membership Results for Agenda Properties

Now that we have identified conditions under which we can guarantee the safety of a given agenda, we want to know how difficult it is to decide whether those conditions are satisfied. As we shall see, this problem is $\Pi_2^p$-complete for each of the classes of aggregation procedures we have considered. $\Pi_2^p$ (also known as coNP$^{\text{NP}}$ or "coNP with an NP oracle") is a complexity class located at the second level of the polynomial hierarchy (Meyer & Stockmeyer, 1972; Stockmeyer, 1976; Arora & Barak, 2009). This is the class of decision problems for which a certificate for a negative answer can be verified in polynomial time by a machine that has access to an oracle for answering queries to SAT (or any other NP-complete problem). To prove a problem $\Pi_2^p$-complete, we have to prove both membership in $\Pi_2^p$ and $\Pi_2^p$-hardness.

We begin by proving membership in $\Pi_2^p$. To do so, we need to provide an algorithm that, when provided with a certificate intended to establish a negative answer, can verify the correctness of that certificate in polynomial time, if we assume that the algorithm has access to a SAT oracle. In the sequel, we shall write MP both for the median property itself and for the problem of deciding whether a given agenda $\Phi$ satisfies the median property, and similarly for the SMP, SSMP and $k$MP.

**Lemma 20.** *MP, SMP, SSMP, and $k$MP are all in $\Pi_2^p$.*

*Proof.* We shall present the proof for $k$MP, which is intuitively the most difficult of the four problems. The proofs for the other three problems are very similar.

We need to give an algorithm that decides the correctness of a certificate for the *violation* of the $k$MP in polynomial time, assuming it has access to a SAT oracle. For a given agenda $\Phi$ (with $m = |\Phi|$), such a certificate is a set $\Delta \subseteq \Phi$ that (a) needs to be inconsistent and that (b) must *not* have any inconsistent subsets of size $\leqslant k$. Inconsistency of $\Delta$ can be checked with a single query to the SAT oracle. If $m' = |\Delta|$, then there are $\sum_{i=1}^{k} \binom{m'}{i}$ nonempty subsets of $\Delta$ of size $\leqslant k$, which is polynomial in $m'$ (and thus also in $m$).[17] Hence, the second condition can be checked by a further polynomial number of queries to the oracle. □

---

17. This figure is not polynomial in $k$, but this does not affect the argument, as $k$ is a constant.

## 5.5 Hardness Results for Agenda Properties

Next, we want to show that MP, SMP, SSMP and $k$MP are all $\Pi_2^p$-hard. This can be done by giving a polynomial-time reduction from a problem that is already known to be $\Pi_2^p$-hard to the problem under investigation. For this purpose, we will make use of *quantified boolean formulas* (QBFs). While QSat, the satisfiability problem[18] for general QBFs, is PSPACE-complete, by imposing suitable syntactic restrictions we can generate complete problems for any level of the polynomial hierarchy. Consider a QBF of the following form:

$$\forall x_1 \cdots x_r \exists y_1 \cdots y_s.\varphi(x_1, \ldots, x_r, y_1, \ldots, y_s)$$

Here $\varphi$ is an arbitrary propositional formula and $\{x_1, \ldots, x_r\} \cup \{y_1, \ldots, y_s\}$ is the set of all propositional variables occurring in $\varphi$ (that is, the above could be any QBF for which any existential quantifiers occur inside the scope of all universal quantifiers). The problem of checking whether a formula of this form is *satisfiable* (i.e., *true*), which we shall denote $\forall\exists$Sat, is known to be $\Pi_2^p$-complete (Arora & Barak, 2009). Below, we shall abbreviate formulas of the above type by writing $\forall \boldsymbol{x} \exists \boldsymbol{y}.\varphi(\boldsymbol{x}, \boldsymbol{y})$.

The basic intuition for why MP and related problems are $\Pi_2^p$-hard is that they share some basic structure with $\forall\exists$Sat, asking a question of the form "for *all* subsets of $\Phi$ that are inconsistent, does there *exist* a subset with a certain property?" Indeed, embedding, say, MP into $\forall\exists$Sat is relatively straightforward. However, here we require the opposite: we need to show that even though $\forall\exists$Sat may appear to be more general than MP and our other agenda problems, it actually can be reduced to each of these problems.

We first prove a technical lemma. Let $\forall\exists$Sat$^2$ be the problem of checking whether a QBF of the following form is true, *given that we already know* that (*i*) $\varphi$ is not a tautology, (*ii*) $\varphi$ is not a contradiction, and (*iii*) $\varphi$ is not logically equivalent to a literal:

$$\forall \boldsymbol{x} \exists \boldsymbol{y}.\varphi(\boldsymbol{x}, \boldsymbol{y}) \,\wedge\, \forall \boldsymbol{x} \exists \boldsymbol{y}.\neg\varphi(\boldsymbol{x}, \boldsymbol{y})$$

**Lemma 21.** $\forall\exists$Sat$^2$ *is* $\Pi_2^p$*-hard.*

*Proof.* By reduction from $\forall\exists$Sat: Given any QBF of the form $\forall \boldsymbol{x} \exists \boldsymbol{y}.\varphi(\boldsymbol{x}, \boldsymbol{y})$, we show that checking its satisfiability is equivalent to running $\forall\exists$Sat$^2$ on $(\varphi \vee a) \wedge b$ with $a$ being universally and $b$ being existentially quantified, for two new propositional variables $a$ and $b$ not occurring in $\varphi$, i.e., to checking the satisfiability of the formula

$$\forall \boldsymbol{x} \forall a \exists \boldsymbol{y} \exists b.[(\varphi(\boldsymbol{x}, \boldsymbol{y}) \vee a) \wedge b] \,\wedge\, \forall \boldsymbol{x} \forall a \exists \boldsymbol{y} \exists b.\neg[(\varphi(\boldsymbol{x}, \boldsymbol{y}) \vee a) \wedge b].$$

First, note that $(\varphi \vee a) \wedge b$ cannot be a tautology, a contradiction, or equivalent to a literal; so the side constraints specified in the definition of $\forall\exists$Sat$^2$ are satisfied. Also note that the first conjunct above is true exactly when the original formula $\forall \boldsymbol{x} \exists \boldsymbol{y}.\varphi(\boldsymbol{x}, \boldsymbol{y})$ is true. This is because $b$ can always be set to true, and the original formula has to be true whenever $a$ is set to false ($a$ falls under the scope of a universal quantifier). Therefore, a positive answer to the $\forall\exists$Sat$^2$ instance above entails a positive answer to the original $\forall\exists$Sat instance. The other direction is immediate, because the second of the above conjuncts is always satisfiable (by making $b$ false). $\square$

---

18. We shall speak of "satisfiability problems" for QBFs, even though strictly speaking for QBFs there is no distinction between satisfiability, truth and validity, as every QBF is a closed formula.

We are now able to prove $\Pi_2^p$-hardness for the SSMP:

**Lemma 22.** SSMP *is* $\Pi_2^p$-*hard*.

*Proof.* We shall give a polynomial-time reduction from $\forall\exists\mathrm{SAT}^2$ to SSMP; the claim then follows from Lemma 21. Take any instance of $\forall\exists\mathrm{SAT}^2$, i.e., the question whether $\forall\boldsymbol{x}\exists\boldsymbol{y}.\varphi(\boldsymbol{x},\boldsymbol{y}) \wedge \forall\boldsymbol{x}\exists\boldsymbol{y}.\neg\varphi(\boldsymbol{x},\boldsymbol{y})$ is true for some $\varphi$ with $\not\models \varphi$, $\varphi \not\models \bot$, and $\not\models \varphi \leftrightarrow \ell$ for literals $\ell$. Suppose $\boldsymbol{x} = \langle x_1, \ldots, x_r \rangle$, and define an agenda as follows:[19]

$$\Phi = \{x_1, \neg x_1, x_2, \neg x_2, \ldots, x_r, \neg x_r, (\varphi \wedge \top), \neg(\varphi \wedge \top)\}$$

We now prove that $\Phi$ violates the SSMP if and only if the answer to our $\forall\exists\mathrm{SAT}^2$-question is NO. To see this, consider the following facts. First, suppose $\Phi$ violates the SSMP. Under what circumstances will this be the case? As $\varphi$ is neither a tautology nor a contradiction, any inconsistent subset of $\Phi$ must be nontrivially inconsistent. Furthermore, by construction of $\Phi$ (consisting largely of literals), any inconsistent subset of $\Phi$ not including a pair of syntactic complements must include either $(\varphi \wedge \top)$ or $\neg(\varphi \wedge \top)$, as well as a (complement-free) subset of $\{x_1, \neg x_1, \ldots, x_r, \neg x_r\}$. That is, the only way of violating the SSMP is to find a subset of literals from $\{x_1, \neg x_1, \ldots, x_r, \neg x_r\}$ to make true that forces either $(\varphi \wedge \top)$ or $\neg(\varphi \wedge \top)$ to be false. But this is precisely the situation in which our instance of $\forall\exists\mathrm{SAT}^2$ requires a negative answer.

For the other direction, suppose the answer to our $\forall\exists\mathrm{SAT}^2$-question is NO. This means that we are able to find an assignment $\rho$ for the variables in $\boldsymbol{x}$ that makes either $\varphi$ or $\neg\varphi$ unsatisfiable. W.l.o.g., suppose we are in the latter situation. Construct a subset of $\Phi$, containing $\neg(\varphi \wedge \top)$, that includes the literal $x_i$ if it is set to true by the assignment $\rho$, and $\neg x_i$ otherwise. This is an inconsistent subset of $\Phi$, and since $\varphi$ is neither a tautology nor a contradiction, this falsifies the SSMP. $\square$

Proving hardness for the SMP works similarly:

**Lemma 23.** SMP *is* $\Pi_2^p$-*hard*.

*Proof.* The construction used is the same as for the proof of Lemma 22. The only additional insight required is the observation that for the special kind of agenda constructed in that proof, the SMP and the SSMP coincide: By excluding formulas $\varphi$ that are equivalent to literals, we ensure that the agenda $\Phi$ constructed in the previous proof does not contain any pairs of logically equivalent formulas. $\square$

For the MP we give a proof using a reduction from the SSMP:

**Lemma 24.** MP *is* $\Pi_2^p$-*hard*.

*Proof.* We will show how to reduce the problem of deciding SSMP to an instance of MP. Let $\Phi$ be an agenda on which we want to test the SSMP and let $\Phi^+ = \{\varphi_1, \ldots, \varphi_m\}$ be the set of non-negated formulas in $\Phi$. Now build the set $\Psi^+$ in the following way: copy all formulas in $\Phi^+$ $m$ times, every time renaming the variables occurring in $\varphi_i$, obtaining the formulas $\varphi_i^j$ for $1 \leqslant i,j \leqslant m$. For every $i$ substitute $\varphi_i^i$ by $\varphi_i^i \vee p^i$, where $p^i$ is a new variable not occurring in any of the $\varphi_i^j$. Finally, add $p^1, \ldots, p^m$ to $\Psi^+$. We obtain the following set:

$$\begin{aligned} \Psi^+ \;=\; & \{p^1, \varphi_1^1 \vee p^1, \ldots, \varphi_m^1, \\ & \phantom{\{} p^2, \varphi_1^2, \varphi_2^2 \vee p^2, \ldots, \varphi_m^2, \\ & \phantom{\{} \vdots \\ & \phantom{\{} p^m, \varphi_1^m, \ldots, \varphi_m^m \vee p^m\} \end{aligned}$$

---

19. Using $(\varphi \wedge \top)$ rather than $\varphi$ ensures that the agenda $\Phi$ does not include doubly-negated formulas.

Define $\Psi = \Psi^+ \cup \{\neg\psi \mid \psi \in \Psi^+\}$. We will now show that $\Phi$ satisfies the SSMP if and only if $\Psi$ satisfies the MP. One direction is immediate. Suppose $\Phi$ does not satisfy the SSMP. Then $\Phi$ must have a mi-subset $\Delta$ of size $k \geqslant 2$.[20] Let $\Delta = \{\varphi_{i_1}, \ldots, \varphi_{i_k}\}$. Then there exists a subset of $\Psi$, namely $\Delta' = \{\neg p^{i_1}, \varphi_{i_1}^{i_1} \vee p^{i_1}, \varphi_{i_2}^{i_1}, \ldots, \varphi_{i_k}^{i_1}\}$, that is a mi-set of size $k+1 \geqslant 3$, thereby falsifying the MP.

For the opposite direction, suppose that $\Psi$ does not satisfy the MP. That is, $\Psi$ has a mi-subset $\Delta$ of size $\geqslant 3$. By construction of $\Psi$, we know that such a subset must only contain formulas with the same superscript or their complements (all other formulas having different variables). If this subset does not contain any $p^i$ or $\neg p^i$, then we can find a copy of it in $\Phi$, which then violates the SSMP, in which case we are done. Clearly, $\Delta$ cannot include both $p^i$ and $\neg p^i$, as that would contradict $|\Delta| \geqslant 3$. So we are left with those cases where $\Delta$ includes either $p^i$ or $\neg p^i$ for some $i$. Then, by minimality, also $\varphi_i^i \vee p^i$ or its negation must be included. We can now reason by cases: (1) if both $p^i$ and $\varphi_i^i \vee p^i$ are in $\Delta$, then by dropping the disjunction we will still get an inconsistent subset, against the assumption of minimality; (2) both $\neg p^i$ and $\neg(\varphi_i^i \vee p^i)$ cannot be in $\Delta$ for the same reason; (3) finally, $p^i$ together with the negation of $\varphi_i^i \vee p^i$ is already inconsistent. Therefore, we can conclude that $\Delta$ must be of the form $\{\neg p^i, \varphi_i^i \vee p^i\} \cup \Delta^i$, where $\Delta^i$ is a set of (one or more) formulas in $\Psi$ with the same superscript $i$. It is now easy to see that the set we obtain when we remove the superscript from $\{\varphi_i^i\} \cup \Delta^i$ is a mi-subset of $\Phi$ that falsifies the SSMP. In particular, $\neg\varphi_i^i \notin \Delta^i$, because $\neg\varphi_i^i \notin \Psi$ by construction, i.e., the mi-subset of $\Phi$ we obtain does not consist of two formulas that are logical complements. □

Finally, we establish hardness for the $k$MP:

**Lemma 25.** $k$MP *is* $\Pi_2^p$*-hard for every* $k \geqslant 2$.

*Proof.* For $k = 2$, the claim has been established by Lemma 24. Now observe that we can use exactly the same construction as in the proof of Lemma 24 to reduce any instance of $k$MP for some $k \geqslant 2$ to an instance of the corresponding $(k{+}1)$MP. Hence, by a simple inductive argument, $k$MP must be $\Pi_2^p$-hard for any finite $k \geqslant 2$. □

## 5.6 Complexity of the Safety of the Agenda

We have shown that deciding whether a given agenda $\Phi$ satisfies the MP, the SMP, the SSMP, or the $k$MP is both in $\Pi_2^p$ and $\Pi_2^p$-hard. Furthermore, in Section 5.3 we have linked these properties to the safety of $\Phi$ for various combinations of axioms. As an immediate corollary to all of these results, we obtain our theorem concerning the complexity of deciding the safety of an agenda:

---

20. The fact that $\Delta$ cannot contain two formulas that are logical complements is not relevant for our proof.

**Theorem 26.** *Deciding the problem of the safety of an agenda is $\Pi_2^p$-complete for any of the following classes of aggregation procedures:*

(i) $\mathcal{F}_\Phi[\mathrm{WR}, \mathrm{A}, \mathrm{S}, \mathrm{M}^\mathrm{I}]$, *consisting only of the majority rule;*
(ii) $\mathcal{F}_\Phi[\mathrm{WR}, \mathrm{A}, \mathrm{U}, \mathrm{S}]$, *the systematic procedures;*
(iii) $\mathcal{F}_\Phi[\mathrm{WR}, \mathrm{A}, \mathrm{U}, \mathrm{N}]$, *the neutral procedures;*
(iv) $\mathcal{F}_\Phi[\mathrm{WR}, \mathrm{A}, \mathrm{U}, \mathrm{I}]$, *the independent procedures;*
(v) *any class of uniform quota rules $F_m$ with $m > n - \frac{n}{k}$ for some $k \geqslant 2$.*

*Proof.* Concerning $\Pi_2^p$-hardness, (i) is a direct consequence of Theorem 11 and Lemma 24. In the same way, (ii) is derived from Theorem 16 and Lemma 23, (iii) from Theorem 17 and Lemma 23, and (iv) from Theorem 18 and Lemma 22. Finally, (v) follows from Theorem 19 together with Lemma 25. Membership in $\Pi_2^p$ follows from Lemma 20 in all five cases. □

That is, not only is it the case that the safety of the agenda can only be guaranteed for structurally simple agendas, but deciding whether a given agenda meets those structural constraints is highly intractable. This is a negative result in the sense that it concerns a problem that we would like to be able to solve efficiently. We should stress that this does not render the problem hopeless. Work on QBF solvers has seen a lot of progress in recent years (see, e.g., Narizzano, Pulina, & Tacchella, 2006), and such tools could be deployed to check whether an agenda satisfies a given type of median property.[21] In any event, understanding how a naturally arising question in JA relates to a difficult but well-studied algorithmic problem such as $\forall\exists\textsc{Sat}$ is interesting and worthwhile in its own right.

## 6. Related Work: Computational Perspectives on Judgment Aggregation

Starting with the work of List and Pettit (2002), most research in JA has focussed either on the philosophical implications of the fact that aggregation may result in an inconsistent outcome or on the derivation of impossibility and characterisation results. The extensive literature in this field has recently been reviewed by List and Puppe (2009). Some work has also explored the links between JA and preference aggregation (Dietrich & List, 2007b; Grossi, 2009; Porello, 2010; Grandi & Endriss, 2011) and several recent contributions have furthermore focussed on the definition and analysis of specific aggregation procedures (Dietrich & List, 2007a; Dietrich & Mongin, 2010; Miller & Osherson, 2009; Lang et al., 2011). Here we shall instead concentrate on contributions to JA that either have a computational slant or that are otherwise relevant to AI.

Besides our own previous work on the subject of the present paper (Endriss et al., 2010a, 2010b), there have been a small number of contributions in computational social choice taking a *computational perspective* on JA (Nehama, 2010; Slavkovik & Jamroga, 2011; Baumeister, Erdélyi, & Rothe, 2011; Baumeister, Erdélyi, Erdélyi, & Rothe, 2012): The first example is the work of Nehama (2010), who proposes a framework for approximate JA in which the goal of finding an aggregation procedure that will never return an inconsistent judgment set is replaced by the goal of finding a procedure under which returning an inconsistent set is highly unlikely. The (negative) result obtained for this framework is that this does however not extend the range of available procedures in a significant way. Second, Slavkovik and Jamroga (2011) extend the standard JA framework with weights (to model differences in influence between individuals) and provide an upper bound on the complexity of the winner determination problem for a family of distance-based aggregation procedures. Third, Baumeister et al. (2011) provide the first study of the computational complexity of the bribery problem in JA, asking whether it is possible to obtain a desired outcome if up to $k$ individual agents can be bribed so as to change their judgment set. Finally, Baumeister et al. (2012) discuss the complexity of various forms of controlling judgment aggregation processes, e.g., influencing the outcome by adding or removing judges.

---

21. As pointed out by one anonymous reviewer, *Answer Set Programming* may also be a useful framework in which to reason about safety problems. The DLV System, for instance, provides a flexible tool for deciding arbitrary problems located at the second level of the polynomial hierarchy (Leone, Pfeifer, Faber, Eiter, Gottlob, Perri, & Scarcello, 2006).

The clearest example for work that explores the integration of ideas from JA with ideas coming from a field traditionally studied in AI is the recent work on connections between JA and *abstract argumentation frameworks* (Rahwan & Tohmé, 2010; Caminada & Pigozzi, 2011): A problem commonly studied in abstract argumentation is how to decide which ones out of a set of arguments that mutually attack each other to accept, which to reject, and on which to remain undecided. Rahwan and Tohmé (2010) study a variant of this problem where a group of agents have to decide which status to award to which argument, a problem that naturally lends itself to be viewed through the lens of JA. In related work, Caminada and Pigozzi (2011) have proposed an approach to JA that involves a translation into an abstract argumentation framework, which makes the tools and techniques of abstract argumentation available to the aggregation of judgments.

A field of research within AI that is closely related to JA is *belief merging* (see, e.g., Konieczny & Pino Pérez, 2002; Maynard-Zhang & Lehmann, 2003; Chopra et al., 2006; Everaere et al., 2007). The work of Konieczny and Pino Pérez (2002), in particular, has inspired the distance-based procedure for JA we have used in this paper. JA and belief merging as modelled by Konieczny and Pino Pérez share interesting features, but ultimately study different problems. While in JA individuals are assumed to submit consistent judgment sets, in belief merging this constraint is enforced only on the outcome. This reflects the view that consistency in belief merging (modelled in terms of an integrity constraint) is a *feasibility* requirement, while in JA it amounts to a *rationality* assumption.

## 7. Conclusions and Future Work

We have studied the computational complexity of three problems in JA: computing the winning judgment set for a given aggregation procedure, deciding whether manipulation would be beneficial for a given agent under a given aggregation procedure and for a given profile, and deciding on the safety of the agenda for a given class of aggregation procedures. We have also proven several safety theorems that link safety to simple structural properties of the agenda and that provide an interesting counterpart to known possibility theorems.

Our results show that, while the winner determination problem is easy for all quota rules and the premise-based procedure, it is intractable for the otherwise attractive distance-based procedure. Regarding strategic manipulation, we have seen that manipulation is NP-hard for the premise-based procedure, which is a positive result. We have also seen that for quota rules the question of manipulation complexity does not arise, at least not for the model of preferences used here. For the distance-based procedure, we have not investigated the complexity of the manipulation problem, because already the winner determination problem was found to be intractable. In our work on the safety of the agenda, we have derived characterisation results for a wide range of procedures, defined in terms of commonly used axioms. We have seen that safety can only be guaranteed for relatively simple agendas and we have also seen that deciding whether these simplicity conditions are met is highly intractable.

While work on the computational aspects of JA has so far been limited to a small number of interesting but scattered contributions, we strongly believe that JA should be taken up as an important research topic in both AI and computational social choice. One important direction to pursue concerns practical algorithms for the problems studied in this paper (as well as for related problems naturally arising in JA). We have already mentioned that existing work on algorithms for the winner determination problem for the Kemeny rule in preference aggregation (Conitzer et al., 2006) may provide a starting point for a working implementation of the distance-based procedure and that work on QBF solvers in automated reasoning (Narizzano et al., 2006) or work on Answer Set Programming (Leone et al., 2006) could prove helpful in tackling the challenges identified by our complexity results regarding the safety of the agenda.

Alongside the development of practical algorithms, improving our understanding of the algorithmic aspects of JA by studying it in the framework of *parameterised complexity* would also be of great interest. In the context of voting, this approach has lead to a number of insightful results (Betzler, 2010). Indeed, for JA, initial steps in this direction have already been taken by Baumeister et al. (2011).

Studying the winner determination problem, both in complexity-theoretic and in practical terms, for the other distance-based procedures proposed by Miller and Osherson (2009) and Lang et al. (2011) also constitutes a very worthwhile direction for future work.

Recall that we have analysed manipulation for one particular way of defining preferences, namely in terms of the Hamming distance to an agent's true set of judgments. Thus, it would be interesting to investigate to what extent changing the definition of manipulation (by altering the notion of induced preference) affects our complexity result. Indeed, other notions of induced preference (and thus manipulation) are conceivable. For instance, a would-be manipulator might only be interested in the status of specific propositions (e.g., the "conclusions") or she might use a different notion of distance, e.g., one of those recently proposed by Duddy and Piggins (2012).

Above we have justified our decision not to study the complexity of the manipulation problem for the distance-based procedure with the fact that already the much more basic winner determination problem is $\Theta_2^p$-complete. An important question that we believe requires discussion in the research community is whether this is indeed a valid argument. In the context of voting, the initial idea of Bartholdi et al. (1989) had been that, say, an NP-hardness result for the manipulation problem for a particular voting rule might suggest that this rule is immune against manipulation in practice. Recent work very strongly suggests that this is not the case (Faliszewski & Procaccia, 2010), and that for the kind of NP-hard problems encountered in this context algorithms that perform well in practice are relatively easy to design (Walsh, 2011). The question now arises whether the same will still be true for hardness results with respect to higher complexity classes. For instance, it is conceivable that it will be possible to design algorithms that can efficiently solve most "typical" instances of the winner determination problem for the distance-based procedure, while it might turn out to be much more difficult to design a similarly successful algorithm for the corresponding manipulation problem. That is, the question arises whether hardness-of-manipulation studies need to be restricted to problems for which winner determination is polynomial, or whether any "jump" in complexity is desirable in principle and might provide some level of protection in practice.

Regarding the safety of the agenda, we have given results for the most natural combinations of axioms that correspond to a weakening of the majority rule, but a similar study could also be conducted for other combinations of axioms. Indeed, it would be interesting to explore how robust our $\Pi_2^p$-completeness results are. That is, an open question that suggests itself is whether there exists an interesting and relevant class of aggregation procedures for which the safety problem falls into a different complexity class.

Generally speaking, we believe that much more work on exploring the obvious potential of JA for AI and multiagent systems is needed. This should lead to both practical advances and the definition of interesting new theoretical problems. Some steps in this direction have recently been taken by Slavkovik (2012), concerning the modelling of collective decision making in multiagent systems, and by Caminada and Pigozzi (2011) and Rahwan and Tohmé (2010), concerning applications of JA to abstract argumentation.

## Acknowledgments

This paper builds on our earlier work on the complexity of judgment aggregation presented at AAMAS-2010 (Endriss et al., 2010a) and COMSOC-2010 (Endriss et al., 2010b). We would like to thank the reviewers and members of the audiences of these meetings, three reviewers for the *Journal of Artificial Intelligence Research*, and the attendees of workshop and seminar talks we have given on this topic at Amsterdam, Barcelona, Chongqing, Luxembourg, Moscow, New Delhi, Padova, Paris, Pisa, and Tilburg for the many helpful suggestions received.

## References

Arora, S., & Barak, B. (2009). *Computational Complexity: A Modern Approach*. Cambridge University Press.

Bartholdi, J. J., Tovey, C. A., & Trick, M. A. (1989). The computational difficulty of manipulating an election. *Social Choice and Welfare*, *6*(3), 227–241.

Baumeister, D., Erdélyi, G., Erdélyi, O. J., & Rothe, J. (2012). Control in judgment aggregation. In *Proceedings of the 6th Starting AI Researchers' Symposium (STAIRS-2012)*. IOS Press.

Baumeister, D., Erdélyi, G., & Rothe, J. (2011). How hard is it to bribe the judges? A study of the complexity of bribery in judgment aggregation. In *Proceedings of the 2nd International Conference on Algorithmic Decision Theory (ADT-2011)*. Springer-Verlag.

Betzler, N. (2010). *A Multivariate Complexity Analysis of Voting Problems*. Ph.D. thesis, University of Jena.

Brandt, F., Conitzer, V., & Endriss, U. (2012). Computational social choice. In Weiss, G. (Ed.), *Multiagent Systems*. MIT Press. In press.

Caminada, M., & Pigozzi, G. (2011). On judgment aggregation in abstract argumentation. *Autonomous Agents and Multi-Agent Systems*, *22*(1), 64–102.

Chevaleyre, Y., Endriss, U., Lang, J., & Maudet, N. (2007). A short introduction to computational social choice. In *Proceedings of the 33rd Conference on Current Trends in Theory and Practice of Computer Science (SOFSEM-2007)*. Springer-Verlag.

Chopra, S., Ghose, A., & Meyer, T. (2006). Social choice theory, belief merging, and strategy-proofness. *Information Fusion*, *7*(1), 61–79.

Conitzer, V., Davenport, A. J., & Kalagnanam, J. (2006). Improved bounds for computing Kemeny rankings. In *Proceedings of the 21st National Conference on Artificial Intelligence (AAAI-2006)*.

Dietrich, F. (2007). A generalised model of judgment aggregation. *Social Choice and Welfare*, *28*(4), 529–565.

Dietrich, F., & List, C. (2007a). Judgment aggregation by quota rules: Majority voting generalized. *Journal of Theoretical Politics*, *19*(4), 391–424.

Dietrich, F., & List, C. (2007b). Arrow's theorem in judgment aggregation. *Social Choice and Welfare*, *29*(1), 19–33.

Dietrich, F., & List, C. (2007c). Strategy-proof judgment aggregation. *Economics and Philosophy*, *23*(3), 269–300.

Dietrich, F., & Mongin, P. (2010). The premiss-based approach to judgment aggregation. *Journal of Economic Theory*, *145*, 562–582.

Dokow, E., & Holzman, R. (2010). Aggregation of binary evaluations. *Journal of Economic Theory*, *145*(2), 495–511.

Duddy, C., & Piggins, A. (2012). A measure of distance between judgment sets. *Social Choice and Welfare*, *39*(4), 855–867.

Endriss, U., Grandi, U., & Porello, D. (2010a). Complexity of judgment aggregation: Safety of the agenda. In *Proceedings of the 9th International Conference on Autonomous Agents and Multiagent Systems (AAMAS-2010)*.

Endriss, U., Grandi, U., & Porello, D. (2010b). Complexity of winner determination and strategic manipulation in judgment aggregation. In *Proceedings of the 3rd International Workshop on Computational Social Choice (COMSOC-2010)*.

Everaere, P., Konieczny, S., & Marquis, P. (2007). The strategy-proofness landscape of merging. *Journal of Artificial Intelligence Research (JAIR)*, *28*(1), 49–105.

Faliszewski, P., & Procaccia, A. D. (2010). AI's war on manipulation: Are we winning?. *AI Magazine*, *31*(4), 53–64.

Gaertner, W. (2006). *A Primer in Social Choice Theory*. LSE Perspectives in Economic Analysis. Oxford University Press.

Grandi, U., & Endriss, U. (2011). Binary aggregation with integrity constraints. In *Proceedings of the 22nd International Joint Conference on Artificial Intelligence (IJCAI-2011)*.

Grossi, D. (2009). Unifying preference and judgment aggregation.. In *Proceedings of the 8th International Conference on Autonomous Agents and Multiagent Systems (AAMAS-2009)*.

Hemachandra, L. A. (1989). The strong exponential hierarchy collapses. *Journal of Computer and System Sciences*, *39*(3), 299–322.

Hemaspaandra, E., Hemaspaandra, L. A., & Rothe, J. (1997). Exact analysis of Dodgson elections: Lewis Carroll's 1876 system is complete for parallel access to NP. *Journal of the ACM*, *44*(6), 806–825.

Hemaspaandra, E., Hemaspaandra, L. A., & Menton, C. (2012). Search versus decision for election manipulation problems. Tech. rep. URCS-TR-2012-971, University of Rochester, Computer Science Department.

Hemaspaandra, E., Spakowski, H., & Vogel, J. (2005). The complexity of Kemeny elections. *Theoretical Computer Science*, *349*(3), 382–391.

Kemeny, J. (1959). Mathematics without numbers. *Daedalus*, *88*(4), 577–591.

Konieczny, S., & Pino Pérez, R. (2002). Merging information under constraints: A logical framework. *Journal of Logic and Computation*, *12*(5), 773–808.

Kornhauser, L. A., & Sager, L. G. (1993). The one and the many: Adjudication in collegial courts. *California Law Review*, *81*(1), 1–59.

Lang, J., Pigozzi, G., Slavkovik, M., & van der Torre, L. (2011). Judgment aggregation rules based on minimization. In *Proceedings of the 13th Conference on Theoretical Aspects of Rationality and Knowledge (TARK-XIII)*.

Leone, N., Pfeifer, G., Faber, W., Eiter, T., Gottlob, G., Perri, S., & Scarcello, F. (2006). The DLV system for knowledge representation and reasoning. *ACM Transactions on Computational Logic*, *7*(3), 499–562.

List, C., & Pettit, P. (2002). Aggregating sets of judgments: An impossibility result. *Economics and Philosophy*, *18*(1), 89–110.

List, C., & Puppe, C. (2009). Judgment aggregation: A survey. In *Handbook of Rational and Social Choice*. Oxford University Press.

Maynard-Zhang, P., & Lehmann, D. J. (2003). Representing and aggregating conflicting beliefs. *Journal of Artificial Intelligence Research (JAIR)*, *19*, 155–203.

Meyer, A. R., & Stockmeyer, L. J. (1972). The equivalence problem for regular expressions with squaring requires exponential space. In *Proceedings of the 13th Annual Symposium on Switching and Automata Theory (SWAT/FOCS-1972)*. IEEE Computer Society.

Miller, M., & Osherson, D. (2009). Methods for distance-based judgment aggregation. *Social Choice and Welfare*, *32*(4), 575–601.

Narizzano, M., Pulina, L., & Tacchella, A. (2006). The QBFEVAL web portal. In *Proceedings of the 10th European Conference on Logics in Artificial Intelligence (JELIA-2006)*. Springer-Verlag.

Nehama, I. (2010). Approximate judgment aggregation. In *Proceedings of the 3rd International Workshop on Computational Social Choice (COMSOC-2010)*.

Nehring, K., & Puppe, C. (2007). The structure of strategy-proof social choice. Part I: General characterization and possibility results on median spaces. *Journal of Economic Theory*, *135*(1), 269–305.

Nehring, K., & Puppe, C. (2010). Abstract Arrowian aggregation. *Journal of Economic Theory*, *145*(2), 467–494.

Papadimitriou, C. H. (1981). On the complexity of integer programming. *Journal of the ACM*, *28*(4), 765–768.

Papadimitriou, C. H. (1994). *Computational Complexity*. Addison Wesley.

Pettit, P. (2001). Deliberative democracy and the discursive dilemma. *Philosophical Issues*, *11*(1), 268–299.

Pigozzi, G. (2006). Belief merging and the discursive dilemma. *Synthese*, *152*(2), 285–298.

Porello, D. (2010). Ranking judgments in Arrow's setting. *Synthese*, *173*(2), 199–210.

Rahwan, I., & Tohmé, F. (2010). Collective argument evaluation as judgement aggregation. In *Proceedings of the 9th International Conference on Autonomous Agents and Multiagent Systems (AAMAS-2010)*.

Rothe, J., Spakowski, H., & Vogel, J. (2003). Exact complexity of the winner problem for Young elections. *Theoretical Computer Science*, *63*(4), 375–386.

Slavkovik, M. (2012). *Judgment Aggregation for Multiagent Systems*. Ph.D. thesis, University of Luxembourg.

Slavkovik, M., & Jamroga, W. (2011). Distance-based judgment aggregation of three-valued judgments with weights. In *Proceedings of the IJCAI-2011 Workshop on Social Choice and Artificial Intelligence*.

Stockmeyer, L. J. (1976). The polynomial-time hierarchy. *Theoretical Computer Science*, *3*(1), 1–22.

Wagner, K. W. (1987). More complicated questions about maxima and minima, and some closures of NP. *Theoretical Computer Science*, *51*(1–2), 53–80.

Walsh, T. (2011). Where are the hard manipulation problems?. *Journal of Artificial Intelligence Research (JAIR)*, *42*, 1–29.